\documentclass[reprint,prl,aps, floatfix]{revtex4-1}
\usepackage{graphicx}
\usepackage{xcolor}
\usepackage{dcolumn}
\usepackage{bm}
\usepackage{amssymb}
\usepackage{braket}
\usepackage{color,amsmath}
\usepackage{comment}
\usepackage{gensymb}
\usepackage{soul,xcolor} 
\setstcolor{red} 
\usepackage{epstopdf}
\usepackage{hhline}
\usepackage{tabularx}
\usepackage{xspace}
\usepackage{layouts}
\usepackage{lineno} 
\usepackage{xr} 

\newcolumntype{C}[1]{>{\centering\arraybackslash}m{#1}}
\newcolumntype{N}{@{}m{0pt}@{}}

\begin{document}

\title{Giant elastoresistance in magic-angle twisted bilayer graphene}

\author{Xuetao Ma$^{1*}$}
\author{Zhaoyu Liu$^{2*}$}
\author{Jiaqi Cai$^{2}$}
\author{Kenji Watanabe$^{3}$}
\author{Takashi Taniguchi$^{4}$}
\author{Xiaodong Xu$^{2,1}$}
\author{Jiun-Haw Chu$^{2\dagger}$}
\author{Matthew Yankowitz$^{2,1\dagger}$}

\affiliation{$^{1}$Department of Materials Science and Engineering, University of Washington, Seattle, Washington, 98195, USA}
\affiliation{$^{2}$Department of Physics, University of Washington, Seattle, Washington, 98195, USA}
\affiliation{$^{3}$Research Center for Electronic and Optical Materials, National Institute for Materials Science, 1-1 Namiki, Tsukuba 305-0044, Japan}
\affiliation{$^{4}$Research Center for Materials Nanoarchitectonics, National Institute for Materials Science, 1-1 Namiki, Tsukuba 305-0044, Japan}
\affiliation{$^{*}$These authors contributed equally to this work.}
\affiliation{$^{\dagger}$jhchu@uw.edu (J.H.C.); myank@uw.edu (M.Y.)}

\begin{abstract}
Strongly correlated and topological phases in moir\'e materials are exquisitely sensitive to lattice geometry at both atomic and superlattice length scales. Twist angle, pressure, and strain directly modify the lattice, and thus act as highly effective tuning parameters. Here we examine electrical transport in twisted bilayer graphene subjected to continuous uniaxial strain. Near the magic angle ($\approx 1.1^{\circ}$), devices exhibit a pronounced elastoresistance that depends on band filling and temperature, with a gauge factor more than two orders of magnitude larger than that of conventional metals. In selected doping regimes the elastoresistance exhibits a Curie–Weiss--like temperature divergence. We discuss possible microscopic origins, including nematic fluctuations and enhanced electronic entropy from fluctuating isospin moments. Our work establishes uniaxial strain as a versatile probe of correlated physics in a moir\'e material.
\end{abstract}

\maketitle
Rapid progress in recent years has led to the prediction and realization of two-dimensional van der Waals (vdW) heterostructures with flat electronic bands~\cite{Balents2020MoireBands,Andrei2020Review,Nuckolls2024Microscopic,Adak2024Berry,Wolf2024Magnetism}. A defining feature of these systems is that specific atomic lattices or moir\'e superlattices suppress electronic kinetic energy near the Fermi level, amplifying the effects of many-body interactions. Such flat bands routinely host symmetry-broken correlated states, unconventional superconductivity, nontrivial topological phases, and other exotic phenomena. Because band flattening is tied to precise lattice geometry, the electronic properties are highly sensitive to external perturbations such as mechanical stress. For magic-angle twisted bilayer graphene (MATBG)~\cite{Cao2018Correlated,Cao2018Superconductivity,Yankowitz2019Tuning} in particular, theory~\cite{Kwan2021KekuleSpiralOrder,Wagner2022PhaseDiagram} and experiment~\cite{Finney2022Magnetotransport,Nuckolls2023QuantumTextures} show that heterostrain introduced during fabrication can stabilize electronic states that are otherwise inaccessible in unstrained devices.

\begin{figure*}[t]
\includegraphics[width=0.85\textwidth]{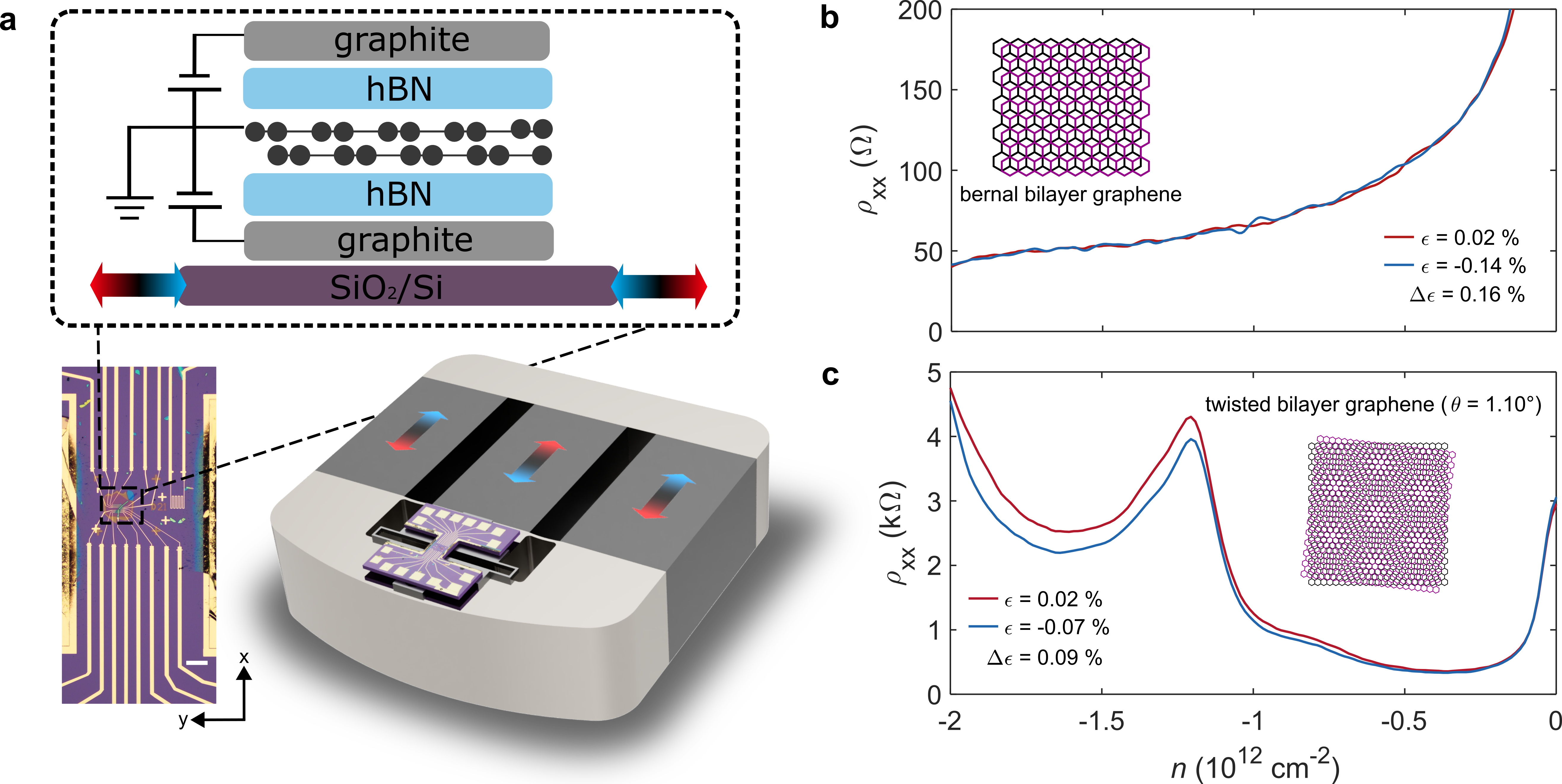} 
\caption{\textbf{Experimental setup and elastoresistance measurements.}
\textbf{a}, (top) Schematic of the dual-graphite gated devices. The active channel layer is either Bernal bilayer graphene or twisted bilayer graphene. The blue (red) arrow pairs indicate compressive (tensile) stress applied to the silicon wafer substrate. (Bottom left) Optical micrograph of the $1.10^{\circ}$ TBG sample. The scale bar is $10~\mu$m. (Bottom right) Drawing of the strain apparatus. Three piezos (dark gray) are epoxied to titanium end-plates (light gray). One titanium plate features a narrow gap region, across which the vdW device resting on a thin silicon wafer can be uniaxially strained upon applying a bias, $V_p$, to the piezos.
\textbf{b}, Measurement of $\rho_{xx}$ versus $n$ for hole-doped BBG (at $D=0$~V/nm) at two values of strain (positive and negative values of $\epsilon$ correspond to tensile and compressive strain, respectively). 
\textbf{c}, The same for MATBG ($\theta=1.10^{\circ}$).
}
\label{fig:1}
\end{figure*}

Despite its importance, strain in moir\'e vdW devices has thus far been unintentional and fixed at the time of sample assembly (with rare exception~\cite{Kapfer2023Programming}). Mapping the properties of a given moir\'e material versus strain therefore requires fabricating many nominally similar devices, each with a different built-in strain profile. This is very challenging, however, due to the microscopic patches of heterogeneous strain---often called “moir\'e disorder’’---that cannot be reproduced precisely from device to device~\cite{Lau2022Reproducibility}. Direct, in-situ control of strain in atomically-thin vdW devices is therefore highly desirable, but remains elusive. 

In strongly correlated bulk crystals, on the other hand, there has been rapid progress over the past 15 years in using uniaxial stress to study the physics of electronic nematicity, unconventional superconductivity, and more~\cite{Chu2012Nematic,hicks2024probing}. Among the various strain techniques that have been developed, the most widely used method for inducing large strain in rigid crystals is a three-piezostack design, in which stress is applied by suspending the sample across a narrow, deformable gap on a piezoelectric platform~\cite{Hicks2014PiezoApparatus}. However, extending this approach to ultra-thin vdW devices, while also maintaining dual-gated geometries and high sample quality, has proven challenging~\cite{cenker2022reversible,hwangbo2024strain}. 

We overcome these hurdles with a recently developed piezoelectric technique that applies continuous uniaxial stress to vdW devices in-situ~\cite{Liu2024StrainControl}. Using this apparatus, we measure the elastoresistance, $\Delta R/R$, of bilayer graphene in both Bernal-stacked and twisted configurations near the magic angle ($\approx 1.1^{\circ}$). Our principal findings are a giant elastoresistance in twisted bilayer graphene (TBG) near the magic angle that far exceeds that of Bernal bilayer graphene (BBG) and most metals, a doping-dependent gauge factor that in some cases mirrors features of the flat-band resistivity, and a gauge factor that generally increases upon cooling, sometimes following Curie–Weiss behavior. Together, these findings establish strain as a new means of probing strongly correlated states in moir\'e materials. 

\medskip\noindent\textbf{Giant elastoresistance in MATBG}

Figure~\ref{fig:1}a depicts our experimental setup, which uses a piezoelectric device to apply continuous uniaxial strain to ultrahigh-quality, dual-gated vdW heterostructures. The apparatus, described in Ref.~\cite{Liu2024StrainControl}, is fully compatible with standard cryogenic electrical transport measurements. Three piezoelectric stacks connected in parallel are driven by a voltage $V_p$; the central stack is poled opposite to the two outer stacks, so differential expansion strains the sample. We note that there is built-in strain in the sample generated during the fabrication of the vdW device, along with additional strain due to differential thermal contraction of the strain cell as it is cooled, such that it is infeasible to determine the true ``zero strain'' state of the the sample. Hence, all the strain values reported in this work are measured relative to $V_p=-20$~V, which corresponds to approximately zero external strain in our setup based on in-situ Raman spectroscopy measurements~\cite{Liu2024StrainControl}. 

A key result is that the elastoresistance of MATBG far exceeds that of Bernal bilayer graphene (BBG). Figures~\ref{fig:1}b-c compare the longitudinal resistivity ($\rho_{xx}$) versus doping ($n$) of BBG and MATBG as a function of the estimated uniaxial strain ($\epsilon$) (see the Supplementary Information for details of calculating $n$ and $\epsilon$). For BBG the resistance remains constant within experimental resolution as $\epsilon$ is varied by $0.16\%$. In contrast, MATBG (with a twist angle of $\theta=1.10^{\circ}$) shows clear strain-induced changes even over a smaller window of $0.09\%$.

\begin{figure*}[t]
\includegraphics[width=\textwidth]{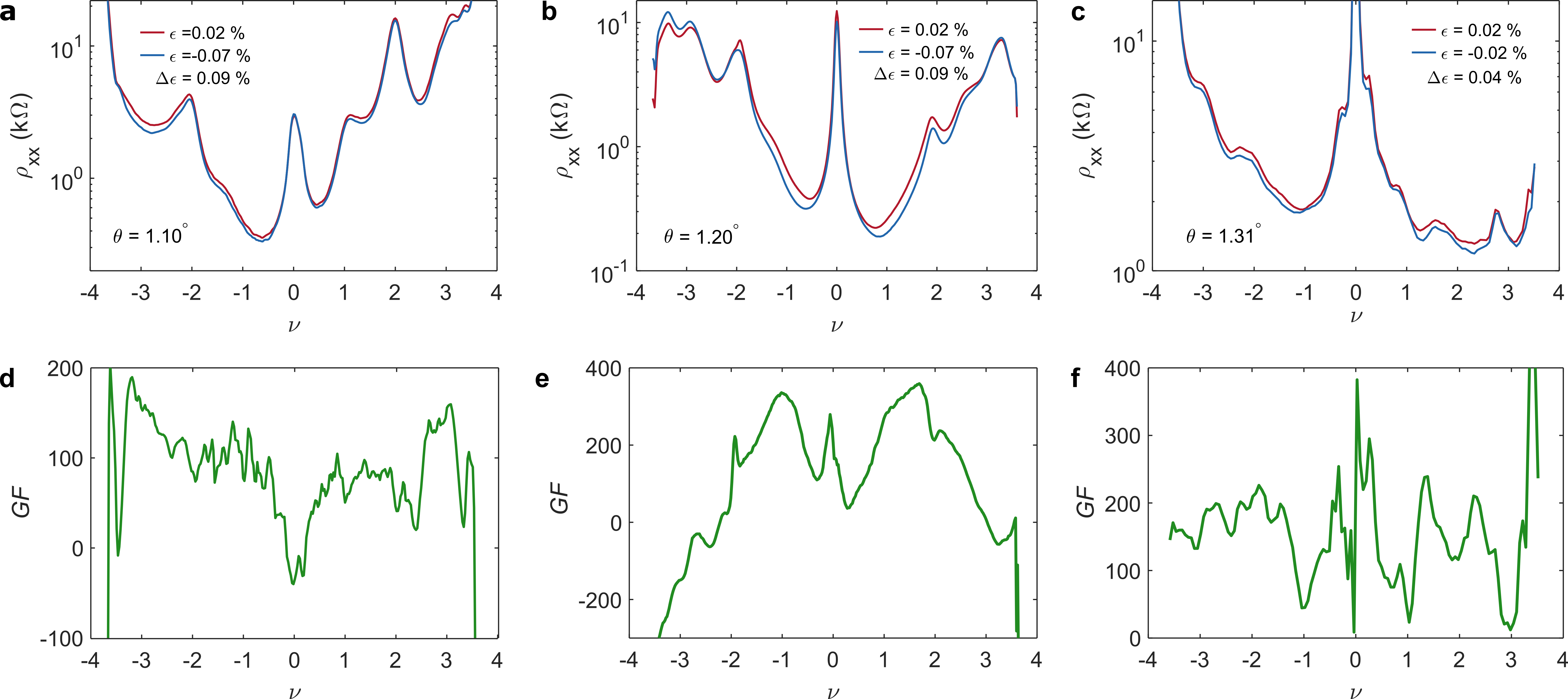} 
\caption{\textbf{Elastoresistance and gauge factor of TBG near the magic angle.}
\textbf{a-c}, Measurement of $\rho_{xx}$ versus $\nu$ at several values of strain for various TBG devices. The twist angles are (\textbf{a}) $\theta=1.10^{\circ}$, (\textbf{b}) $\theta=1.20^{\circ}$, and (\textbf{c}) $\theta=1.31^{\circ}$. All measurements are taken at $T=2$~K. 
\textbf{d-f}, Calculated $GF$ versus $\nu$ for each device in \textbf{a-c}, respectively. 
}
\label{fig:2}
\end{figure*}

Figure~\ref{fig:2}a shows an expanded measurement of $\rho_{xx}$ in the MATBG device across the entire flat band (with $n$ converted to the band filling factor $\nu$ as described in the Supplementary Information). Compressive strain, corresponding to $\epsilon<0$, monotonically decreases the resistance across most of the flat band. Comparable behavior is observed in additional TBG devices with twist angles of $1.20^{\circ}$ and $1.31^{\circ}$, shown in Figs.~\ref{fig:2}b-c, respectively. These two devices may also be unintentionally aligned with one of the encapsulating hexagonal boron nitride (hBN) flakes, as analyzed in Supplementary Information Fig.~\ref{fig:bn_alignment}.

Crucially, the elastoresistance is approximately linear with $V_p$ in most measurements (Supplementary Information Fig.~\ref{fig:ER_Vp}). This enables a faithful calculation of the gauge factor---the relevant figure of merit for strain measurements, $GF=(\Delta R/R)/\Delta\epsilon$---by converting $V_p$ to $\epsilon$. The conversion procedure is based on electrical measurements of several calibrated strain gauges in our apparatus, along with estimates for the strain transmission from the silicon wafer to the graphene and other related quantities (Supplementary Information Fig.~\ref{fig:strain_conversion}). In the Supplementary Information, we discuss the relevant considerations for extracting the gauge factor from our measurements, and argue that our conversion provides a lower bound on its true value.

\begin{figure}[t]
\includegraphics[width=3.4in]{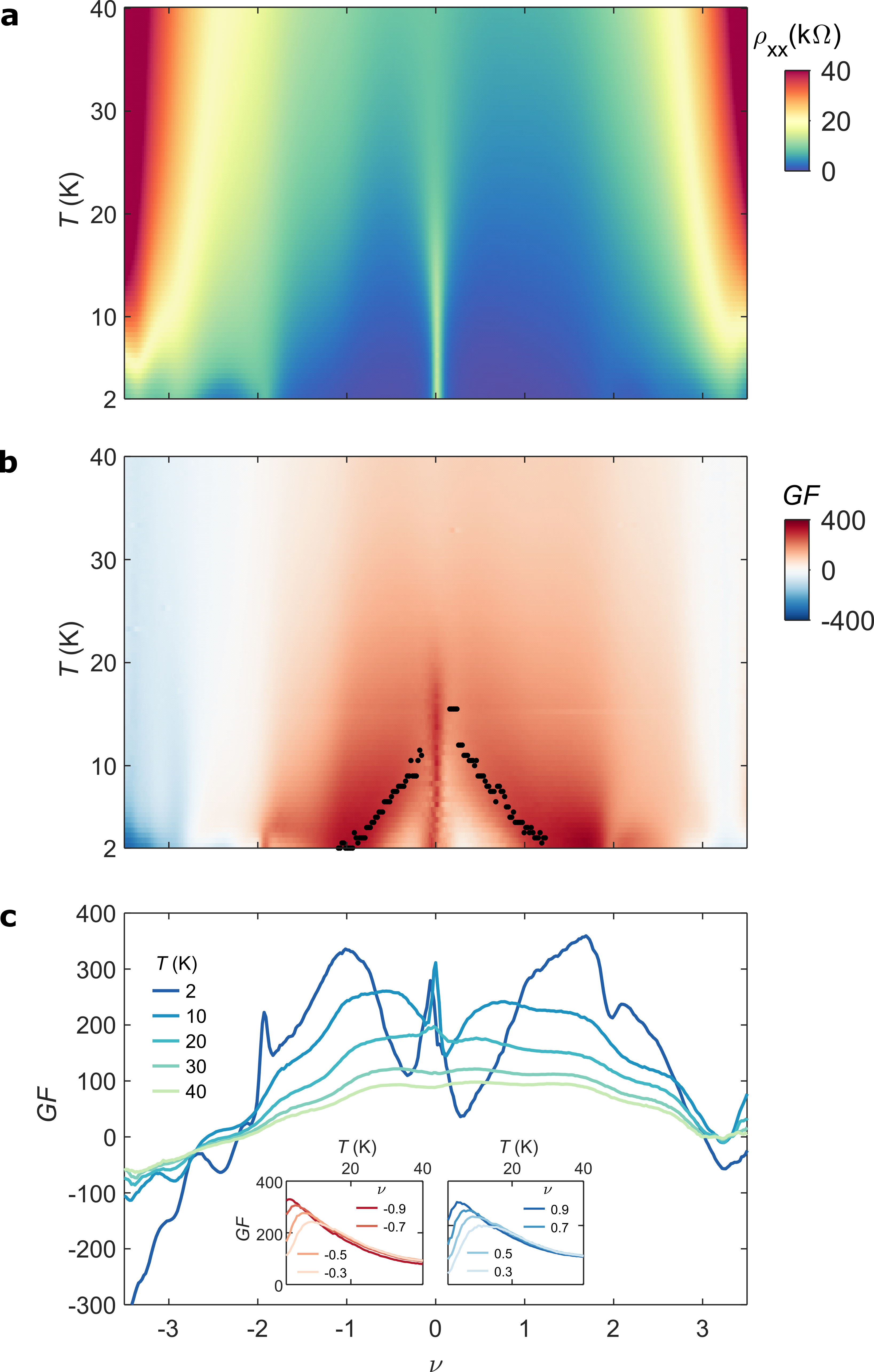} 
\caption{\textbf{Temperature dependence of the resistance and elastoresistance in TBG.}
\textbf{a}, Measurement of $\rho_{xx}$ versus $\nu$ and $T$ in the $\theta=1.20^{\circ}$ TBG device.
\textbf{b}, Map of the $GF$ versus $\nu$ and $T$. Black markers denote the maximum value of $GF$ between $-1.1 < \nu < -0.15$ and $0.15 < \nu < 1.25$.
\textbf{c}, Line traces of $GF$ versus $\nu$ for several fixed $T$. (inset) $GF$ versus $T$ for selected values of $\nu$.
}
\label{fig:3}
\end{figure}

The gauge factor can be rewritten as:
\[
GF=\frac{\Delta R/R}{\Delta L/L}=\frac{\Delta R/R}{\Delta\epsilon} = 1 + 2\nu_P + \frac{\Delta\rho/\rho}{\epsilon}.
\]
with $\nu_P$ the Poisson ratio, which separates the geometric ($1+2\nu_P$) and electronic ($\Delta\rho/\rho)/\epsilon$ contributions. The gauge factor for ordinary metals is typically of order unity. In most metals $\Delta\rho/\rho$ is negligible, and the gauge factor is thus almost entirely geometric (with typical Poisson ratio of $\nu_P \approx 0.1 - 0.5$). Notably, however, in some lightly-doped semiconductors such as silicon~\cite{Hynecek1974}, and strongly correlated materials near nematic transitions~\cite{Chu2012Nematic}, the electronic term dominates and greatly enhances $GF$.

Figures~\ref{fig:2}d-f plot $GF$ for the three TBG devices. All exhibit very large gauge factor values, in some cases approaching $400$, two orders of magnitude above those of conventional metals. The gauge factor varies sharply with carrier density and in some cases shows features aligned with integer $\nu$. Although the elastoresistance is usually positive, rapid variations---including sign reversals---appear, most prominently for $-4<\nu<-2$ in the $1.20^{\circ}$ device (Fig.~\ref{fig:2}e). As discussed later, a measured gauge factor of order $10^{2}$ points to a dominant electronic contribution.

\medskip\noindent\textbf{Temperature dependence of the elastoresistance}

We next examine the temperature dependence of elastoresistance in TBG, which offers key insights into correlation effects at elevated temperatures. Figure~\ref{fig:3}a shows $\rho_{xx}$ within the flat band of the $\theta=1.20^{\circ}$ TBG device up to $T=40$~K. The data closely match earlier reports from devices near the magic angle, notably including resistance bumps at several integer $\nu$ that persist to the highest measured temperature~\cite{Polshyn2019LinearT}. 

We extract the gauge factor by first measuring $\rho_{xx}$ versus $\nu$ as $T$ is swept at fixed $V_p$, then repeat similar measurements for several values of $V_p$. Figure~\ref{fig:3}b shows the resulting map of the temperature-dependent $GF$ across the flat band. Two features stand out. First, there are vertical structures corresponding to step-like changes in the gauge factor that roughly align with integer $\nu$, mirroring the resistance bumps in Fig.~\ref{fig:3}a. Second, $GF$ traces a dome-like profile: it is generally largest near the charge neutrality point, especially at high $T$, and smaller or even negative near the band edges. Line cuts of $GF(\nu)$ at selected temperatures (Fig.~\ref{fig:3}c) confirm these trends. The dome with steps at integer $\nu$ resembles thermodynamic measurements of electronic entropy in MATBG~\cite{Rozen2021Pomeranchuk,Saito2021IsoPomeranchuk,Zhang2025HeavyTunneling}, suggesting a possible connection that we will discuss in more detail later.

\begin{figure*}[t]
\includegraphics[width=\textwidth]
{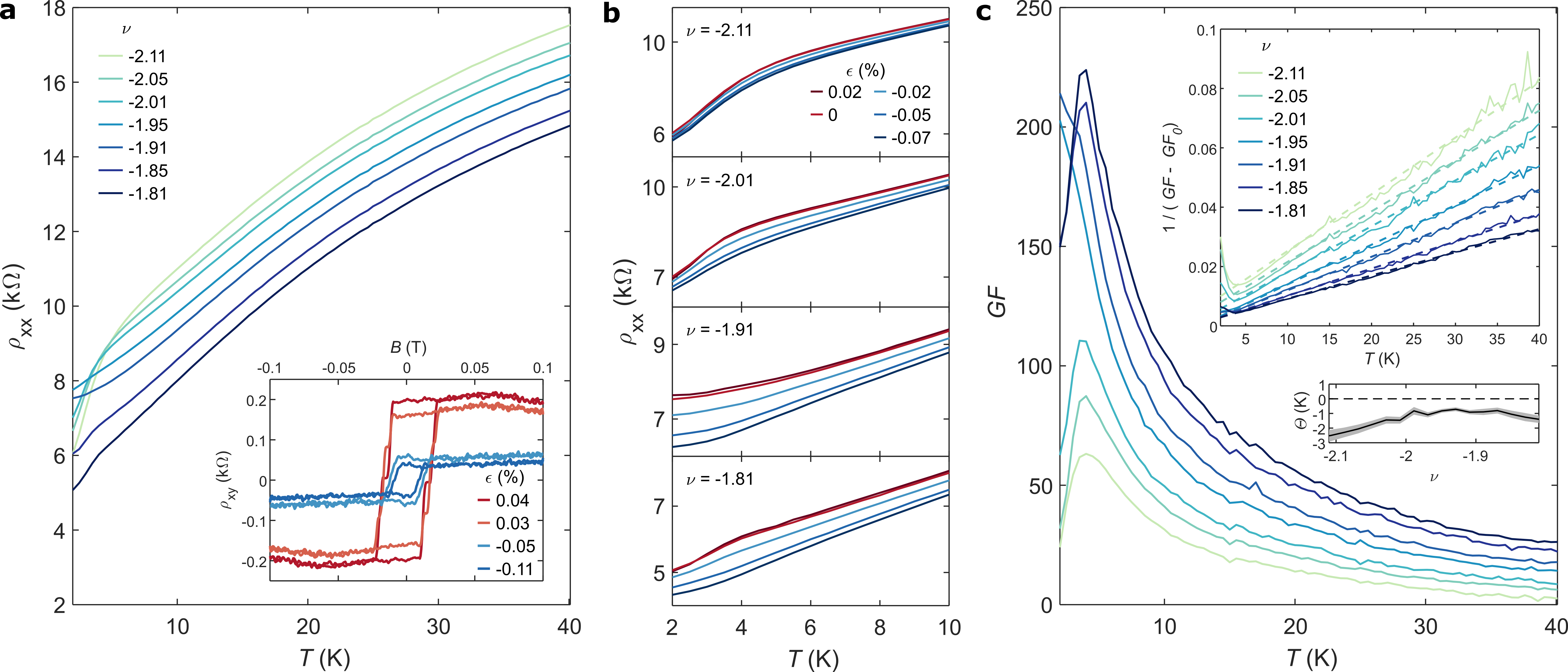} 
\caption{\textbf{AHE and Curie-Weiss law behavior of $GF$ near half filling.}
\textbf{a}, Line traces of $\rho_{xx}$ versus $T$ at several values of $\nu$ near $-2$ in the TBG device with $\theta=1.20^{\circ}$. (inset) Antisymmetrized anomalous Hall effect at $\nu = -1.99$ for several $\epsilon$.
\textbf{b}, $\rho_{xx}$ versus $T$ for several $\epsilon$ at various fixed $\nu$.  
\textbf{c}, $GF$ versus $T$ for the same values of $\nu$ as in \textbf{(a)}. (top inset) Plots of the same data as $1/(GF-GF_0)$ versus $T$, with dashed lines showing a linear fit based on the Curie-Weiss law. (bottom inset) Fitted value of $\Theta$ versus $\nu$. The shaded area indicates the error bar, defined by a 95\% confident interval.} 
\label{fig:4}
\end{figure*}

\medskip\noindent\textbf{Strain-tunable anomalous Hall effect}

Figure~\ref{fig:4}a shows $\rho_{xx}(T)$ for several values of $\epsilon$ near half-filling of the moir\'e valence band ($\nu=-2$). This filling is adjacent to the most robust superconducting pocket in many MATBG devices~\cite{Balents2020MoireBands}. Numerous studies have attributed the corresponding normal state to intervalley-coherent order, yet the present device instead shows an anomalous Hall effect (AHE) near $\nu = -2$ (inset of Fig.~\ref{fig:4}a) and no superconductivity down to millikelvin temperatures. Our prior analysis on this sample indicates that the AHE likely reflects valley-polarized order driven by staggered sublattice potentials from a nearly aligned hBN substrate~\cite{Tseng2022AHEHalfFilling}. Despite this unconventional ground state, $\rho_{xx}(T)$ above 2~K closely follows typical MATBG behavior: a large resistivity that rises rapidly for $4 \lesssim T \lesssim 25$~K, and an abrupt change at temperatures below a few kelvin~\cite{Polshyn2019LinearT,Cao2020StrangeMetal,Jaoui2022QuantumCritical}.

The inset of Fig.~\ref{fig:4}a plots the Hall resistance ($\rho_{xy}$) as an external magnetic field ($B$) is swept back and forth for different $\epsilon$. Each trace shows a hysteresis loop centered at $B=0$, the hallmark signature of the AHE. Small steps in $\rho_{xy}$ appear just below the coercive field; these likely correspond to Barkhausen jumps caused by magnetic domain-wall motion. Similar features are common in orbital-magnetic states of moir\'e materials, and have previously been described as a ``Chern mosaic'' formed by a dense patchwork of microscopic domains with alternating signs of the magnetization~\cite{Grover2022ChernMosaic}.

We find that straining the sample towards compressive systematically reduces the AHE amplitude by more than a factor of two over the accessible strain window, and slightly reduces the coercive field. Curiously, the AHE sign can be switched in an irreversible fashion by sweeping $V_p$ at fixed $\nu$ and $B\approx0$ (Supplementary Fig.~\ref{fig:AHE_strain}). Such a strain response likely arises from a microscopic reconfiguration of orbital-magnetic domains, in addition to any intrinsic coupling of strain to the band structure. Notably, the data were taken very near a Barkhausen jump in $\rho_{xy}$, where the magnetic domains walls are likely to have instabilities in their locations. Straining the system and returning it to the original configuration could potentially lead to a change in the location of the domain walls, corresponding to the sign flip in $\rho_{xy}$.

\medskip\noindent\textbf{Curie-Weiss law behavior near $\nu=-2$}

We now focus in detail on the elastoresistance near $\nu=-2$. Figure~\ref{fig:4}b shows $\rho_{xx}(T)$ at different $\epsilon$ for several $\nu$ near half filling. For curves with an abrupt downturn at low temperature, the resistivity is most varied as a function of strain near the ``knee'' at $T \approx 3-4$~K. As the temperature is further lowered, the curves converge back towards one another. In contrast, for the curves at $\nu=-1.91$ where there is a low-temperature saturation of $\rho_{xx}$ rather than an abrupt drop, the curves instead continue to separate down to the base temperature of $2$~K.

This connection can be seen more clearly from the extracted gauge factors as a function of $T$ (Fig.~\ref{fig:4}c). In all, the gauge factor rises sharply as the temperature is lowered from $40$~K. For the values of $\nu$ in which $\rho_{xx}(T)$ exhibits a sharp downturn below a few kelvin, the corresponding gauge factor reaches a maximum and then also turns down at low $T$. For $\nu=-1.91$ in particular, where there is instead a low-$T$ saturation in $\rho_{xx}$, the gauge factor continues to rise down to $2$~K.

The shape of $GF(T)$ closely follows a Curie–Weiss form
\[
GF = \frac{C}{T - \Theta} + GF_0,
\]
where $C$ is the Curie constant, $\Theta$ is the Weiss temperature, and $GF_0$ is a temperature-independent offset. The inset of Fig.~\ref{fig:4}c plots $(GF - GF_0)^{-1}$ versus $T$; the resulting linearity over a decade change of $T$ confirms the Curie–Weiss law behavior.

Curie–Weiss scaling, however, is not universal. Supplementary Information Figure~\ref{fig:fullGF_120} presents $GF(T)$ at other fillings for this device. Some traces follow the same $1/T$ power law, whereas others do not. In the magic-angle device with $\theta=1.10^{\circ}$, we find no clear Curie–Weiss regime at any $\nu$ (Supplementary Information Figs.~\ref{fig:fullGF_110} and~\ref{fig:GFtemp_110}), although this device had the most non-linear evolution of elastoresistance with $V_p$ (Supplementary Information Fig.~\ref{fig:ER_Vp}). Despite this, most $GF(T)$ curves in both devices share the qualitative feature of an initial increase with cooling; a subset also show a low-temperature maximum followed by a downturn toward base temperature.

\medskip\noindent\textbf{Discussion and outlook}

A key open experimental question is the microscopic nature of the strain pattern produced by a given piezo voltage $V_p$: does the applied stress generate homostrain (symmetric deformation of both graphene sheets) or heterostrain (asymmetric deformation)? The two couple very differently to the electronic structure, with homostrain resulting in a negligible change to the flat bands but heterostrain greatly perturbing them~\cite{Wang2023StrainTransport}. Despite this, they are difficult to distinguish using simple transport measurements alone, as they each result in only tiny changes to the moir\'e unit cell area and thus immeasurably small changes in the doping required to fill the flat bands (see Supplementary Information). 

We speculate that our experiment may predominately induce homostrain, since the entire vdW device is plasma etched into a mesa and pinned down with evaporated metal contacts, and also because there is large interlayer friction between the two graphene sheets near the magic angle~\cite{Kapfer2023Programming} that may further serve to pin them together. However, the applied uniaxial stress could also introduce a heterostrain component, and even a small amount would help explain the large gauge factors we observe. Future experiments that can directly probe the microscopic structure of the sample will be needed to unambiguously identify the nature of the induced strain.

Another challenge is to separate the effect of isotropic (in-plane area-changing) strain, $\epsilon_{\mathrm{iso}} = (\epsilon_{xx} + \epsilon_{yy})/2$, and anisotropic (shape-changing) strain, $\epsilon_{\mathrm{aniso}} = (\epsilon_{xx} - \epsilon_{yy})/2$ (where $\epsilon_{ii}$ denotes the strain along the $i$-direction). These two strain components are both present upon the application of uniaxial stress. Ideal TBG possesses an in-plane three-fold rotational symmetry ($C_3$), such that a $120^{\circ}$ rotation maps each carbon atom onto an equivalent atom in the same layer and simultaneously carries the moir\'e lattice (and its Brillouin zone) onto itself. The $K_{m}$ and $K'_{m}$ corners of the hexagonal moir\'e Brillouin zone are therefore symmetry-related, and the Dirac points are pinned there. $\epsilon_{\mathrm{aniso}}$ breaks the three‑fold rotational symmetry ($C_{3}$), leaving no remaining in‑plane rotational symmetry. As a result, the moir\'e lattice becomes oblong, its Brillouin zone distorts, and the Dirac points shift away from the original $K_{m}$ and $K'_{m}$ positions~\cite{Bi2019FlatBands}. In practice, a small heterostrain ($\sim 0.1\%$) is already present in typical MATBG devices after assembly~\cite{Lau2022Reproducibility}, breaking the $C_3$ symmetry even before external stress is applied. Whether the built-in heterostrain or the externally applied uniaxial strain dominates the electronic structure depends on their relative magnitudes and orientations, an issue that remains open for the devices in our study.

We now discuss possible microscopic origins of our observations and outline future directions. Near the magic angle, the electronic dispersion of TBG is extremely sensitive to the moir\'e wavelength and to the atomic arrangement within the moir\'e unit cell. Calculations that include lattice relaxations already show sizable band structure changes compared with rigid-lattice models~\cite{Nam2017LatticeRelax}, and modest heterostrain can further split van Hove singularities and create open Fermi surfaces~\cite{Wang2023StrainTransport}. Because both twist angle and heterostrain are known to affect the resistance~\cite{Polshyn2019LinearT,Cao2020StrangeMetal,Jaoui2022QuantumCritical,Wang2023StrainTransport}, it is natural to attribute the large elastoresistance to strain-induced band structure changes, possibly further modified by changes in the electron-phonon coupling (see additional discussion in the Supplementary Information). 

We specifically consider two mechanisms that can potentially explain the doping- and temperature-dependence of the gauge factor we observe: scattering from fluctuating isospin moments and coupling to nematic fluctuations. The first is captured by heavy-fermion--inspired models, in which flat-band orbitals localized at AA stacking sites hybridize with itinerant carriers comprising the dispersive portion of the band~\cite{Song2022HeavyFermion}. Such a picture is broadly consistent with related theories of MATBG~\cite{Po2019FragileTopo,Carr2019WannierTB,Calderon2020EightOrbital,Kang2021Cascades,Shi2022HeavyRep,Datta2023HeavyQuasiparticles}, which are generally compatible with emerging experimental evidence for coexisting local moments and itinerant carriers based on transport, compressibility, and thermoelectric measurements up to $\sim 100$~K~\cite{Rozen2021Pomeranchuk,Saito2021IsoPomeranchuk,LuqueMerino2024HeavyThermo,Ghosh2025Thermopower,Zhang2025HeavyTunneling}. The localized isospin moments carry large electronic entropy, and inelastic scattering of itinerant electrons off these moments can potentially account for the large and strongly temperature-dependent resistivity of MATBG. 

It has been shown previously that $\epsilon_{\mathrm{iso}}$ couples to the electronic entropy of an iron-based superconductor, giving rise to large and diverging temperature-dependent elastoresistance~\cite{Wiecki2020Elastoresistance,wiecki2021emerging}. In our MATBG devices the gauge factor similarly increases with decreasing temperature and then rolls over upon cooling to base temperature. The black markers in Fig.~\ref{fig:3}b track the temperature of the maximum value of $GF$ found between $-1.1 < \nu < -0.15$ and $0.15 < \nu < 1.25$, with a selection of corresponding $GF(T)$ curves shown in the inset of Fig.~\ref{fig:3}c. Notably, prior measurements of entropy in MATBG show that it grows most quickly with temperature around charge neutrality~\cite{Rozen2021Pomeranchuk,Saito2021IsoPomeranchuk,Zhang2025HeavyTunneling}, consistent with a transition between a regime of fluctuating local moments at high temperature and a Fermi liquid at low temperature. The turnover in $GF(T)$ may act as a sensitive indicator of this crossover.

A second possibility is that coupling of $\epsilon_{\mathrm{aniso}}$ to nematic ordering leads to a diverging elastoresistance. Angle-resolved transport on magic-angle twisted trilayer graphene shows electronic anisotropy across the entire flat band that persists to $\sim 100$~K~\cite{Zhang2025Interplay}. Although comparable data are not yet available for MATBG, the similarity of their flat bands suggests that nematic fluctuations in MATBG~\cite{Jiang2019ChargeOrder,Fernandes2020NematicTwist,Cao2021Nematicity} could be coupled to $\epsilon_{\mathrm{aniso}}$ and thus contribute to the elastoresistance. Notably, the Curie-Weiss dependence is most robust near $\nu=-2$ (see Supplementary Information Fig.~\ref{fig:CW_fit}), extending over a wide range of temperature and with a corresponding Weiss temperature near zero (lower inset of Fig.~\ref{fig:4}c). Similar elastoresistance behavior has previously been attributed to a nematic quantum critical point in iron-based superconductors~\cite{Chu2012Nematic,Worasaran2021NematicQC}, and may be linked to the linear-in-temperature scaling of its resistance~\cite{Lederer2017NematicQCP}. An unusual feature in the TBG sample, however, is that the region of vanishing $\Theta$ appears to be spread across a broad range of $\nu$ rather than limited to a single point.

Within our current experimental limits, the temperature dependence of the gauge factor could reflect a combination of these two mechanisms. Disentangling them will require measuring all the in-plane resistivity tensor components ($\rho_{xx}$, $\rho_{yy}$ and $\rho_{xy}$), enabling a distinction between isotropic and anisotropic elastoresistivity coefficients. The “sunflower’’ device geometry~\cite{Wu2017RotSymmetry,Zhang2024Nonreciprocity}, which allows current to flow both parallel and perpendicular to the stress axis, is well suited for such measurements. 

Looking forward, we note that our strain platform is compatible with local probes such as scanning SQUID magnetometry, single-electron transistors, and scanning tunneling microscopy, enabling real-space imaging of strain-induced lattice distortions and a more direct determination of their influence on the correlated states in MATBG. Finally, this approach can be easily extended to other flat-band moir\'e and crystalline vdW heterostructures, where tunable strain may shed light on superconductivity, nematic and crystalline electronic orders, and integer- or fractional-quantized anomalous Hall states.

\section*{Acknowledgments}
This work, including the development of the strain cell technique and the study of strain-tunable elastoresistance, was primarily supported by the University of Washington Molecular Engineering Materials Center, a U.S. National Science Foundation Materials Research Science and Engineering Center (DMR-2308979). The development of the graphene devices was supported by the Army Research Office under award number W911NF-25-1-0012 to M.Y. X.X., J.H.C., and M.Y. acknowledge support from the State of Washington-funded Clean Energy Institute. J.-H.C. also acknowledges support from the David and Lucile Packard Foundation and the Gordon and Betty Moore Foundation’s EPiQS Initiative, grant GBMF6759. K.W. and T.T. acknowledge support from the JSPS KAKENHI (Grant Numbers 21H05233 and 23H02052) and World Premier International Research Center Initiative (WPI), MEXT, Japan. This work made use of shared fabrication facilities at UW provided by NSF MRSEC 2308979.

\section{Author Contributions} 
X.M. and Z.L. built the strain cell apparatus, fabricated the samples, and performed the strain-dependent transport measurements and data analysis. J.C. assisted with the experimental setup under the supervision of X.X. K.W. and T.T. provided the hBN crystals. J.H.C. and M.Y. conceived and supervised the study, and wrote the manuscript with input from all authors.

\section*{Competing interests}
The authors declare no competing interests.

\section*{Additional Information}
Correspondence and requests for materials should be addressed to Jiun-Haw Chu or Matthew Yankowitz.  

\section*{Data Availability}
Source data are available for this paper. All other data that support the findings of this study are available from the corresponding author upon request.

\bibliographystyle{naturemag}
\bibliography{references}

\newpage

\renewcommand{\figurename}{Supplementary Information Fig.}
\renewcommand{\thesubsection}{S\arabic{subsection}}
\setcounter{secnumdepth}{2}
\setcounter{figure}{0} 
\setcounter{equation}{0}

\onecolumngrid
\newpage
\section*{Supplementary Information}

\textbf{Device fabrication.} 
Mechanically exfoliated monolayer and Bernal bilayer graphene flakes were identified using optical microscopy. For TBG devices, the monolayer graphene was separated into multiple pieces using polymer-free anodic oxidation nanolithography~\cite{Li2018ElectrodeFree}. The vdW heterostructures were assembled using a standard dry-transfer technique with a polycarbonate (PC)/polydimethylsiloxane (PDMS) stamp~\cite{Wang2013OneDim}. The devices were constructed by sequentially picking up flakes of: graphite, hBN, TBG (or BBG), hBN, graphite (except for the $\theta=1.1^{\circ}$ TBG device, in which there was no top graphite). For TBG samples, the graphene flakes were rotationally misaligned by a small angle by rotating the stage after picking up the first monolayer graphene.

Separately, a $50~\mu$m-thick doped silicon wafer with a 285~nm capping layer of SiO$_2$ was cut using a laser to create a central bridge region along with two outer support bridges. The thin wafer was affixed to a thicker substrate wafer using PMMA as a temporary glue. The completed vdW stack was then dropped onto the central bridge region of the thin wafer. We used standard electron beam lithography and CHF$_3$/O$_2$ plasma etching to define vdW stacks into a Hall bar geometry and standard metal deposition techniques (Cr/Au)~\cite{Wang2013OneDim} to make electrical contact. Optical micrographs of the four samples studied in this work are shown in Supplementary Information Fig.~\ref{fig:micrographs}.

Finally, the thin wafer was detached from the substrate wafer by dissolving the PMMA glue in acetone. The wafer was affixed to a metal sample mount using Stycast epoxy and screwed onto the strain cell. The outer supporting bridges of the thin silicon wafer were broken with tweezers, and wires were hand-pasted to establish electrical connection between the vdW device and custom printed circuit boards mounted on the strain cell. The strain cell was then wired onto an insert for low-temperature measurements. The entire sample fabrication and strain cell preparation procedure is described in detail in Ref.~\onlinecite{Liu2024StrainControl}. 

\textbf{Transport measurements.} 
All devices have graphite top and bottom gates, except for the $\theta=1.10^{\circ}$ TBG device which only has a bottom gate. For the dual-gated TBG devices, transport measurements were performed by sweeping one gate voltage with the other gate held at ground. For the $\theta=1.20^{\circ}$ TBG device, the back gate voltage ($B_{bg}$) was swept, whereas for the $1.31^{\circ}$ device it was the top gate ($V_{tg}$). For the BBG device, both gates were swept simultaneously in order to separate $n$ from the out-of-plane electric displacement field, $D$. In all cases, $n$ and $D$ were defined as $n= \left(C_{\text{bg}} V_{\text{bg}}+C_{\text{tg}} V_{\text{tg}}\right) / e$ and $D=\left(C_{\text{tg}} V_{\text{tg}} - C_{\text{bg}} V_{\text{bg}}\right) / 2 \epsilon_0$, where $C_{\text{tg}}$ and $C_{\text{bg}}$ are the top and bottom gate capacitance per unit area, $e$ is the elementary charge and $\epsilon_0$ is the vacuum permittivity. $C_{\rm{tg}}$ and $C_{\rm{bg}}$ were estimated by a standard procedure of fitting the slopes of integer quantum Hall states in Landau fan measurements (Supplementary Information Figs.~\ref{fig:fans} and~\ref{fig:BBG_full}).

Electrical transport measurements were performed in a PPMS Dynacool with a base temperature of $2$~K and in a Bluefors dilution refrigerator operated down to a nominal base temperature of $100$~mK as measured by a thermometer on the strain cell. Voltage was applied to the silicon gate to reduce the contact resistance. We convert measured longitudinal resistance ($R$) to $\rho_{xx}$ as $\rho_{xx}=(W/L)R$, where $W$ is the width of the Hall bar and $L$ is the distance between the centers of the voltage probes. We assume the aspect ratio of the Hall bar remains fixed as a function of strain, and thus adopt a single conversion between $R$ and $\rho_{xx}$ for all $V_p$. This is an approximation since in reality the aspect ratio may change slightly as stress is applied, however this geometric correction is very small compared with the large gauge factor values of order $10^2$ we measure, and can thus be safely ignored. The measurement of $\rho_{xx}$ is oriented along the direction of uniaxial stress in all devices, except for the $\theta=1.10^{\circ}$ TBG device in which is oriented at a $45$-degree angle. We do not know the relative orientation of the strain axis with the moir\'e or atomic lattices. The optical micrographs in Supplementary Information Fig.~\ref{fig:micrographs} denote the contacts used as the source (S) and drain (D) in each measurement. Uniaxial stress is applied along the $x$-direction. In all devices, measurements of $\rho_{xx}$ ($\rho_{xy}$) are taken with the voltage probes labeled A and B (A and C). 

\textbf{Twist angle determination.} 
To determine the twist angle ($\theta$) of the TBG devices, we first extracted the carrier density, $n_{s}$, required to fully fill the four-fold--degenerate moir\'e valence and conduction bands (i.e., the value of $n$ at which the insulating states at full band filling occur). This value is further refined by fitting the quantum Hall states appearing in a magnetic field (Supplementary Information Fig.~\ref{fig:fans}). The filling factor was then defined according to $\nu = n/(n_{s}/4)$. We calculated the twist angle using $n_{s}$ = 8$\theta^{2} / \sqrt{3} a^{2}$, where $a=0.246$~nm is the lattice constant of graphene. For the $\theta=1.10^{\circ}$ and $1.20^{\circ}$ TBG devices, in which we measured several pairs of voltage probes, we found a small twist angle inhomogeneity of approximately $0.01^{\circ}$. 

\textbf{Possible alignment with hBN.}
The TBG devices with $\theta=1.20^{\circ}$ and $1.31^{\circ}$ may also be closely rotationally aligned with one of the hBN layers. There are two pieces of evidence for such alignment. The first comes from inspecting optical micrographs of the completed stacks, before they were processed into devices (Supplementary Information Figs.~\ref{fig:bn_alignment}a-b). The red and blue dashed lines denote the straight edges of the two hBN flakes, which typically have either zig-zag or armchair termination and form modulo-$30^{\degree}$ angles. The white dashed line denotes the straight edge of one of the graphene layers. Notably, in both devices the graphene edge is very closely aligned with top hBN edge, modulo $30$~degrees (see the second dashed white line). This indicates the likelihood of alignment either close to $0^{\circ}$ or $30^{\circ}$, with the former corresponding to an additional long-wavelength moir\'e pattern between the graphene and hBN. However, we cannot distinguish between these two possibilities from optical micrographs alone, since we do not know whether a given edge in each flake is zig-zag or armchair.

A second means of assessing alignment to hBN is to check for an energy gap at the charge neutrality point (CNP). TBG devices typically only exhibit such a gap when closely aligned with hBN. Supplementary Information Figures~\ref{fig:bn_alignment}c-d show measurements of $\rho_{xx}$ versus $T$ at the CNP for both devices, exhibiting a sharp increase in resistivity at low temperatures. The insets show the same data in an Arrhenius plot. The CNP exhibits a narrow range of temperature with hints of thermally activated behavior. We extract a small band gap, $\Delta_g$, from slope of the linear fit (black dashed line) using $\rho \propto e^{\frac{\Delta}{2 k_B T}}$, where $k_B$ is the Boltzmann constant. We caution, however, that the total change in resistance is relatively small, much less than one decade in the $1.20^{\circ}$ device in particular. Thus, this analysis provides only indirect evidence for the possibility of hBN alignment in these two devices.  

\textbf{Strain calibration.} 
Our experimental setup generally has two different types of strain gauges. The first is a commercial silicon or metal foil gauge mounted onto the central of the three piezos in the strain cell. The gauge factor is pre-calibrated, such that the strain induced in the piezostack can be extracted by performing a two-terminal measurement of the resistance of the strain gauge as $V_p$ is swept. Separately, we evaporate a meandering gold pattern onto the Si/SiO$_2$ wafer nearby the vdW device. By measuring the four-terminal resistivity of this gold wire and comparing to a prior calibration curve from Ref.~\onlinecite{Liu2024StrainControl}, we can separately estimate the amount of strain induced directly into the Si/SiO$_2$ wafer substrate. The strain transmitted from the cell to the wafer can be estimated using the ratio of these measurements, and was found to be between 0.15 and 0.19 for the three devices in which we have both measurements (the BBG device, and the $\theta=1.10^{\circ}$ and $1.20^{\circ}$ TBG devices). Due to problems during metal liftoff, the evaporated on-chip gold strain gauge was inoperable in the $\theta=1.31^{\circ}$ TBG device, so in this case we assumed a strain transmission equal to the average of the other three devices (0.176).

Our strain technique has several possible sources of systematic error in converting from the bias applied to the piezos ($V_p$) to the uniaxial strain induced in the graphene channel ($\epsilon$). First, as noted in the main text, we do not have a direct measure of the ``zero-strain'' point in our setup, as our strain gauges are only sensitive to relative changes in strain. In order to report values of $\epsilon$, we thus assume that $V_p=-20$~V corresponds to $\epsilon=0$ for all samples, as determined from cryogenic Raman spectroscopy in one prior device~\cite{Liu2024StrainControl}.

Second, each evaporated gold meander on SiO$_2$ has a slightly different gauge factor. We have previously established a calibration to estimate the gauge factor, $GF_{Au}$, based on the measured resistivity of the evaporated gauge $\rho_{Au}$~\cite{Liu2024StrainControl}. However, there is a small uncertainty introduced in this conversion. Supplementary Information Fig.~\ref{fig:strain_conversion} shows the conversion procedure from $V_p$ to $\epsilon$ for each device in this study, based on the raw measured $\rho_{Au}$ for each device, along with corresponding resistance measurements of the commercial silicon and foil gauges ($R_{Si}$ and $R_{foil}$).

Third, when calculating the gauge factor we assume a linear relationship between $\Delta \rho/\rho$ and $\epsilon$ (equivalently, that there is a linear evolution of $R$ with $V_p$). Representative curves at different $n$ for each device are shown in Supplementary Information Fig.~\ref{fig:ER_Vp}. The assumption of linearity is well-justified over much or all of the accessible strain range for the BBG device and the $\theta=1.20^{\circ}$ and $1.31^{\circ}$ TBG devices. However, we see more substantial deviations from linearity for the $\theta=1.10^{\circ}$ TBG device. It is not clear whether this nonlinearity is due to poor strain transmission, or reflects physics beyond the basic assumptions we make in this work.

Fourth, and likely most substantially, we do not have a direct measurement of the amount of strain transmitted from the Si wafer to the graphene channel. The strain transmission cannot exceed unity, but may be substantially smaller. To be conservative, we assume perfect strain transmission from the wafer to the graphene channel, which necessarily overestimates $\epsilon$ and thus underestimates the gauge factor, since the two are inversely related. Therefore, all reported values of $GF$ in this work should be treated as lower bounds.

\textbf{Assessment of the Curie-Weiss fitting.}
We define several criteria that must be simultaneously met in order to establish the applicability of the Curie-Weiss law to $GF(T)$ measurements. First, the $R^2$ coefficient should be very close to $1$ (typically larger than $0.99$), indicative of a faithful fit to the model. Second, the fitted offset parameter, $GF_0$, should be close to zero. This parameter represents a temperature-independent offset to the gauge factor, and must be small in any physically realistic scenario. Third, the fit should be performed over at least a decade change in temperature.

Supplementary Information Figures~\ref{fig:CW_fit}a-b show $R^2$ and $GF_0$ for the best fit to the data shown in Fig.~3b of the main text (see also Supplementary Information Fig.~\ref{fig:fullGF_120}). The three criteria are met simultaneously for a small range of doping surrounding $\nu=-2$. There appears to be another narrow region of doping around $\nu=1.5$ that may also satisfy these criteria, although $GF_0$ varies rapidly with $\nu$ in this case. Supplementary Information Figures~\ref{fig:CW_fit}c-d show the corresponding fitted values of $\Theta$ and $C$. Future work is needed to better understand the origin and extent of Curie-Weiss law dependence.

\textbf{Moir\'e unit cell area with homo- and hetero-strain.} Here, we derive expressions for the change in the area of the moir\'e unit cell---and thus the density needed to fully fill the lowest moir\'e bands---for both uniaxial homostrain and heterostrain. We start with the equation for the real-space area of the moir\'e unit cell ($A_{\mathrm m}$) for a generic hexagonal bilayer, as derived in Ref.~\onlinecite{Kogl2023MoireStraintronics}:

\[
  A_{\mathrm m}
  =\frac{a_1 a_2|\sin\beta|}{\,|\Delta|}\,(1+\delta)^2\,c_1\,c_\mu,
\]
with
\[
\begin{aligned}
  c_1 &= (1+\epsilon_c)^2-\epsilon_s^{\,2}, \\[4pt]
  c_\mu &= (1+\mu\epsilon_c)^2-\mu^{2}\epsilon_s^{\,2},\\[4pt]
  \Delta &= c_\mu(1+\delta)^2 + c_1
           - 2(1+\delta)\Bigl[(1+\epsilon_c+\mu\epsilon_c)
                +\mu(\epsilon_c^{2}-\epsilon_s^{\,2})\Bigr]\cos\theta.
\end{aligned}
\]
Here, $a_1$ and $a_2$ are the primitive lattice vectors of the top layer, $\beta$ is the angle between primitive lattice vectors, \(\delta\) is the lattice mismatch between the two layers, \(\theta\) is the interlayer twist angle, \(\bm\epsilon\) is the in‑plane strain tensor of the bottom layer, \(\epsilon_c=(\epsilon_{xx}+\epsilon_{yy})/2=\epsilon_{iso}\) and  
\(\epsilon_s=\bigl[(\epsilon_{xx}-\epsilon_{yy})^{2}/4+\epsilon_{xy}^{2}\bigr]^{1/2}=\bigl[\epsilon_{aniso}^{2}+\epsilon_{xy}^{2}\bigr]^{1/2}\). The parameter \(\mu\) describes the strain transfer between the two layers (\(\mu=0\) for heterostrain and \(\mu=1\) for homostrain). 

For twisted bilayer graphene, we set \(\delta=0\), \(a_1=a_2=a=0.246\;\mathrm{nm}\), and \(\beta=60^{\circ}\). Thus, the area of the undistorted monolayer graphene unit cell is $A_g=a_1a_2|\sin\beta|=\sqrt{3}a^{2}/2$. For the case of uniaxial strain along the zig‑zag (\(x\)) direction, with $\epsilon_{xy}=0$, the strain tensor becomes:

\[
  \bm\epsilon=
  \begin{pmatrix}\epsilon_u & 0\\[2pt] 0 & -\nu_P\epsilon_u\end{pmatrix},
\]
where $\epsilon_u$ is the uniaxial strain ($\nu_P\approx0.165$ for graphene).

To second order in \(\epsilon_u\), this yields
\[
c_1 = 1+(1-\nu_P)\epsilon_u-\nu_P\epsilon_u^{2},\quad
c_\mu = 1+\mu(1-\nu_P)\epsilon_u-\mu^{2}\nu_P\epsilon_u^{2}.
\]

Inserting \(c_1\) and \(c_\mu\) into \(\Delta\) and keeping terms only up to
\(O(\theta^{2},\epsilon_u^{2})\), we find:

\[
\,
\Delta \approx \theta^{2}+(2\mu\,-\mu\,^2-1)\nu_P\,\epsilon_u^{2}\,.
\]

For the case of uniaxial heterostrain ($\mu=0$):

\[
  A_{\mathrm m}^{(\text{het})}
  =\frac{A_g}{\bigl|\theta^{2}-\nu_P\epsilon_u^{2}\bigr|}[1+(1-\nu_P)\epsilon_u-\nu\epsilon_u^2]\approx\frac{A_g}{\bigl|\theta^{2}-\nu_P\epsilon_u^{2}\bigr|}.
\]

For uniaxial homostrain ($\mu=1$):

\[
  A_{\mathrm m}^{(\text{homo})}
  \approx \frac{A_g}{\theta^{2}}[1+2(1-\nu_P)\epsilon_u].
\]

Denoting the unstrained moir\'e unit cell area as $A_0=A_g/\theta^{2}$, and the change in the moir\'e unit cell area as $\Delta A_{\mathrm m} = A_{\mathrm m}^{(\text{strain})}-A_0$, in the limit of small strain we find:

\[
\frac{\Delta A_{\mathrm m}^{(\text{het})}}{A_0}
   \approx \frac{\nu_P\,\epsilon_u^{2}}{\theta^{2}}
   ,\qquad
\frac{\Delta A_{\mathrm m}^{(\text{homo})}}{A_0}
   \approx 2(1-\nu_P)\epsilon_u.
\]

In graphene, each moir\'e unit cell accommodates four electronic states (spin and valley). The density that corresponds to full filling of a moir\'e miniband in the absence of strain is $n_{s0} = \frac{4}{A_{\text{0}}}$. Under heterostrain, the density corresponding to full filling becomes
\[
  n_s^{(\text{het})}
      \approx\frac{4}{A_g}\,
       \bigl|\theta^{2}-\nu_P\epsilon_u^{2}\bigr|.
\]
For homostrain,
\[
  n_s^{(\text{homo})}
      \approx\frac{4}{A_g}\,
        \frac{\theta^{2}}
             {1+2(1-\nu_P)\epsilon_u}.
\]

Denoting the change in the doping required to fill the moir\'e bands as $\Delta n_s = n_{s}^{(\text{strain})}-n_{s0}$, we find:

\[
  \frac{\Delta n_s^{(\text{het})}}{n_{s0}}
  \approx -\frac{\nu_P\,\epsilon_u^{2}}{\theta^{2}}
  ,\qquad
  \frac{\Delta n_s^{(\text{homo})}}{n_{s0}}
  \approx-\frac{2(1-\nu_P)\epsilon_u}{1+2(1-\nu_P)\epsilon_u}.
\]

For values of the twist angle and uniaxial strain relevant to our experiment, the changes in the doping required to fully fill the bands are too small to measure for both homostrain and heterostrain. Consider, for example, the case of a freestanding TBG with $\theta=1.10^{\circ}$ and a change in uniaxial strain of $0.1$\%. The equations above predict that the moir\'e unit cell area changes by $\approx0.045$\% for heterostrain and $\approx0.17$\% for homostrain. This corresponds to $|\Delta n_s^{(\text{het})}|\approx1.3\times10^{9}$~cm$^{-2}$ and $|\Delta n_s^{(\text{homo})}|\approx4.7\times10^{9}$~cm$^{-2}$. Both of these values are small compared with the uncertainty in fitting $n_s$ using the projection of quantum oscillations from Landau fan diagrams, and also comparable to the typical scale of charge disorder in dual graphite-gated devices ($\approx 5\times10^{9}$~cm$^{-2}$). Thus, it is not possible within our experimental resolution to distinguish between homostrain and heterostrain from simple transport measurements alone.

To illustrate this, Supplementary Information Fig.~\ref{fig:fans_strain} shows an example of Landau fans taken in the $\theta=1.20^{\circ}$ TBG device for two different values of strain differing by $\epsilon=0.15$\%. Within experimental resolution, all of the quantum oscillation features appear in identical locations within the fan diagram, reflecting our inability to detect strain-induced changes in the moir\'e unit cell area in this way.

\textbf{Strain tuning of the anomalous Hall effect.} 
We consider in more detail the strain-tunable behavior of the AHE in the $\theta=1.20^{\circ}$ TBG device near $\nu=-2$. Supplementary Information Figure~\ref{fig:AHE_strain}a shows a color map of the amplitude of the AHE, $\Delta\rho_{\mathrm{AHE}}$, defined as the difference between the maximum and minimum values of $\rho_{xy}$ in the the anomalous Hall loop. We see that $\Delta\rho_{\mathrm{AHE}}$ is largest around $\nu=-1.99$, and grows continuously as $\epsilon$ is made more positive. Supplementary Information Figure~\ref{fig:AHE_strain}b shows temperature-dependent measurements of $\Delta\rho_{\mathrm{AHE}}$ for different values of $\epsilon$. In all cases, the AHE onsets just below $5$~K, and generally achieves larger values for more positive $\epsilon$. The continuous evolution of the AHE amplitude suggests that some portion of the strain tunability is an intrinsic effect, potentially related to the coupling of strain to the energy splitting between valleys generating the underlying orbital magnetism~\cite{Tseng2022AHEHalfFilling}.

However, there are also signatures of strain coupling to disorder-related features of the AHE. Supplementary Information Figures~\ref{fig:AHE_strain}c and d show measurements of $\rho_{xy}$ as the strain is continuously swept from one extreme value to the other and back again. Both measurements are taken at $B$ extremely close to zero and $\nu=-2.01$, but in the former the sweep is started at $\epsilon=-0.11$\%, whereas in the latter it is started at $\epsilon=0.04$\%. The corresponding starting points are indicated by black markers atop the corresponding $\rho_{xy}$ loops shown in the insets. The $\rho_{xy}$ curves trace each other essentially perfectly when starting from $\epsilon=-0.11$\%, but exhibit an irreversible jump when starting from $\epsilon=0.04$\%. We speculate that this may be related to the motion of orbital magnetic domain walls, since in the latter case the starting point is very near a Barkhausen jump, whereas in the former it is not. 

\textbf{Possible role of electron-phonon coupling.}
In addition to the microscopic mechanisms considered in the main text, electron-phonon scattering may also play an important role in determining the elastoresistance of TBG. Recent ARPES~\cite{Chen2024EPC} and quantum twisting microscope~\cite{Birkbeck2025QTM} studies reveal strong electron–phonon coupling in TBG, mediated by a low-energy “moir\'e phonon’’ or phason mode corresponding to antisymmetric interlayer shear~\cite{Koshino2019MoirePhonons,Ochoa2019PhasonCoupling}. Because atomic displacements are greatly magnified at the moir\'e scale~\cite{Wu2018PhononTBG,Lian2019PhononSuperconductor}, strain couples efficiently to this mode and can markedly affect resistance. A detailed characterization of the strain field in our devices, particularly whether it is homo- or heterostrain, will be essential for assessing the phonon contribution.

\newpage

\begin{figure*}[h]
  \centering
  \includegraphics[width=0.9\textwidth]{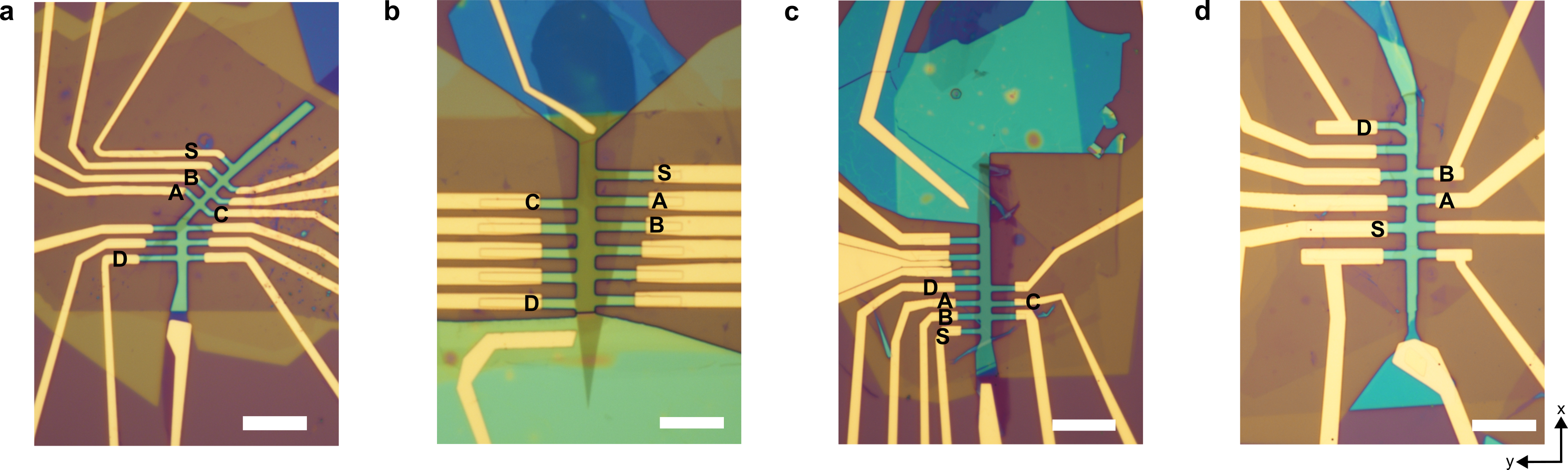}
  \caption{\textbf{Optical micrographs of the devices.}
  \textbf{a}, TBG device with $\theta=1.10^{\circ}$;
  \textbf{b}, TBG device with $\theta=1.20^{\circ}$;
  \textbf{c}, TBG device with $\theta=1.31^{\circ}$;
  \textbf{d}, Bernal bilayer graphene device. In all images, the source and drain pins used during measurements are marked as S and D, respectively. $\rho_{xx}$ is measured using pins A and B, and $\rho_{xy}$is measured using pins A and C. Uniaxial stress is applied along the $x$-direction, as indicated by the axes on the bottom right. The scale bars are $10~\mu$m.}
  \label{fig:micrographs}
\end{figure*}

\begin{figure*}[h]
  \centering
  \includegraphics[width=\textwidth]{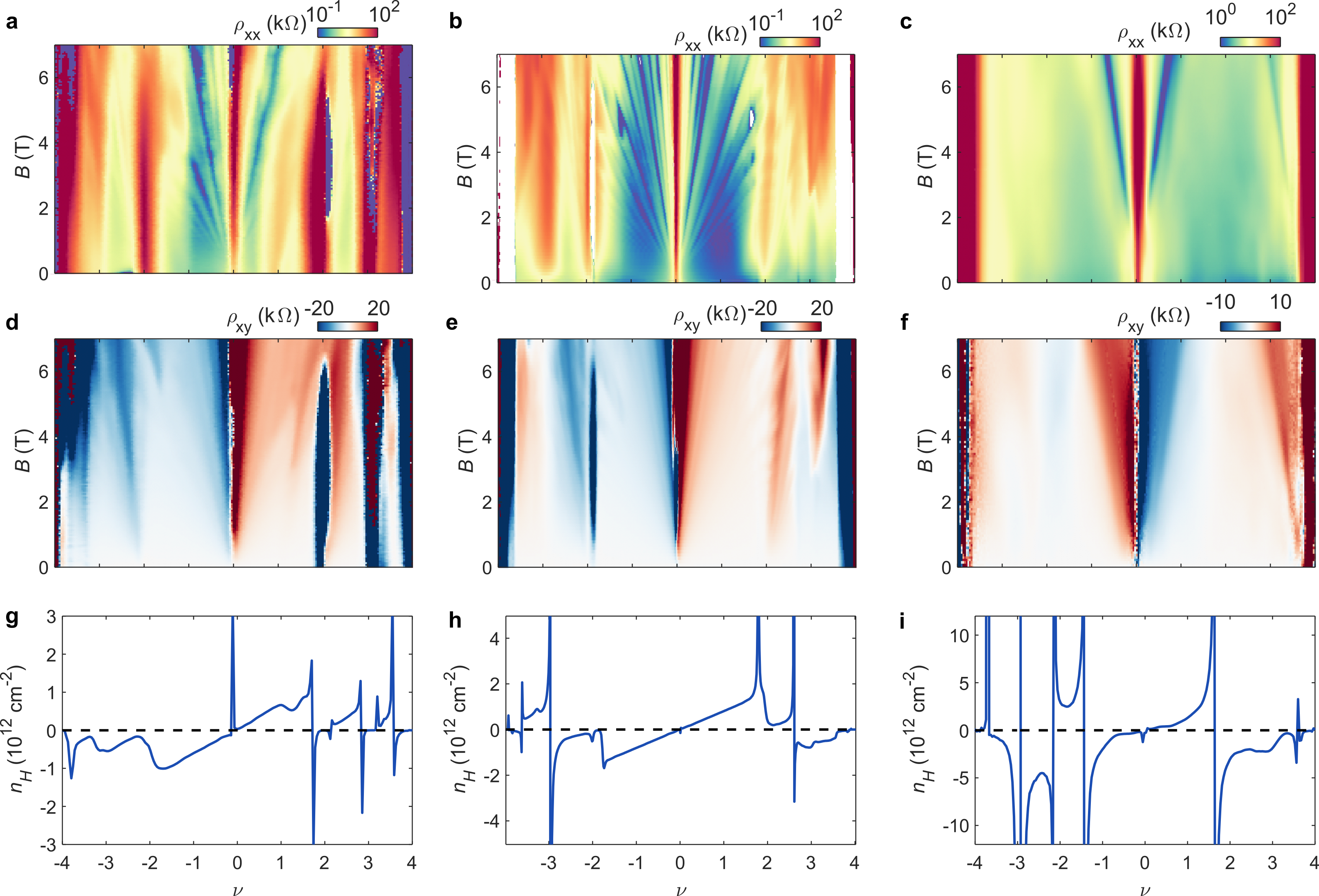}
  \caption{\textbf{Landau fan diagrams from the TBG devices.}
  \textbf{a-c}, Landau fans of longitudinal resistivity, $\rho_{xx}$, from the TBG devices with \textbf{a}, $\theta=1.10^{\circ}$ taken at $V_p=0$ and $T=100$~mK; \textbf{b}, $\theta=1.20^{\circ}$ taken at $V_p=0$ and $T=100$~mK; \textbf{c} $\theta=1.31^{\circ}$ taken at $V_p=-80$~V and $T=2$~K.
  \textbf{d-f}, The same fans of Hall resistance, $\rho_{xy}$. The fans from the $\theta=1.31^{\circ}$ TBG device are field symmetrized/antisymmetrized.
  \textbf{g-i}, Hall density, $n_H=1/(e R_{xy})$, extracted by fitting $\rho_{xy}$ up to 2 T.}
  \label{fig:fans}
\end{figure*}

\begin{figure*}[h]
  \centering
  \includegraphics[width=0.9\textwidth]{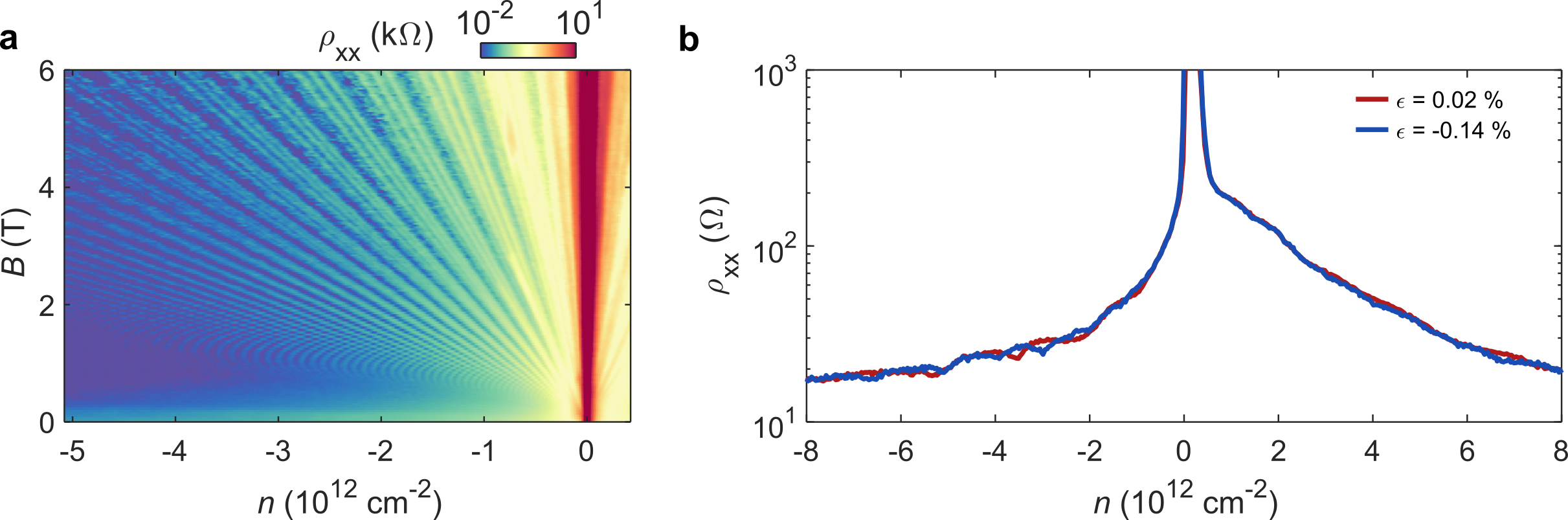}
  \caption{\textbf{Additional characterization of Bernal bilayer graphene.}
  \textbf{a}, Landau fan diagram of the longitudinal resistivity, $\rho_{xx}$, taken at $D=0$ with $V_p=-5$~V and $T=100$~mK.
  \textbf{b}, Measurements at $\rho_{xx}$ at two values of strain, taken at $B=0$. The shoulder on electron-doped side ($n>0$) likely arises due to unintentional doping at the contacts.}
  \label{fig:BBG_full}
\end{figure*}

\begin{figure*}[h]
  \centering
  \includegraphics[width=0.9\textwidth]{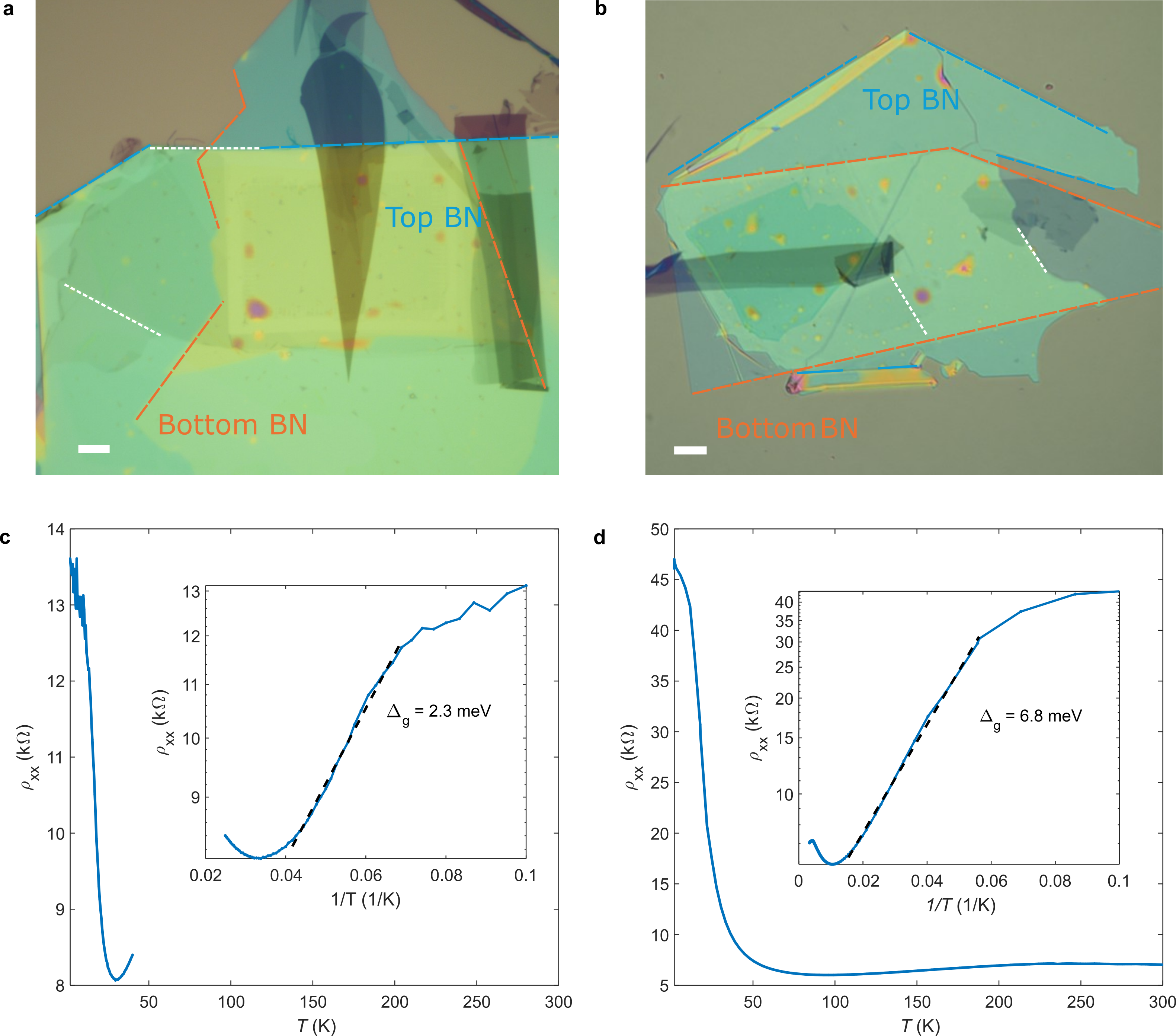}
  \caption{\textbf{Possible alignment of tBLG and hBN in the TBG devices with $\theta=1.20^{\circ}$ and $1.31^{\circ}$.}
  Optical micrographs of the stacks of TBG devices with \textbf{a}, $\theta=1.20^{\circ}$ and \textbf{b}, $\theta=1.31^{\circ}$. Selected straight edges of the top and bottom hBN flakes are denoted by dashed blue and orange lines. A graphene edge is denoted by the white dashed line. The second white dashed line is drawn atop an hBN edge either parallel to the graphene or at a $30^{\circ}$ angle, indicating the potential for close rotational alignment between the two (modulo $30^{\circ}$). The scale bars are 5 $\mu$m.
  \textbf{c}, Measurement of $\rho_{xx}$ versus $T$ at the charge neutrality point ($\nu=0$) for the $\theta=1.20^{\circ}$ TBG device. The inset shows the same data on an Arrhenius plot, with a fitted activation energy of $\Delta_g=2.3$~meV.
  \textbf{d}, The same for the $\theta=1.31^{\circ}$ TBG device, with $\Delta_g=6.8$~meV. In both cases, $\Delta_g$ is potentially an overestimate of the true CNP gap owing to the small range of thermally activated behavior in the device.}
  \label{fig:bn_alignment}
\end{figure*}

\begin{figure*}[h]
  \centering
  \includegraphics[width=0.9\textwidth]{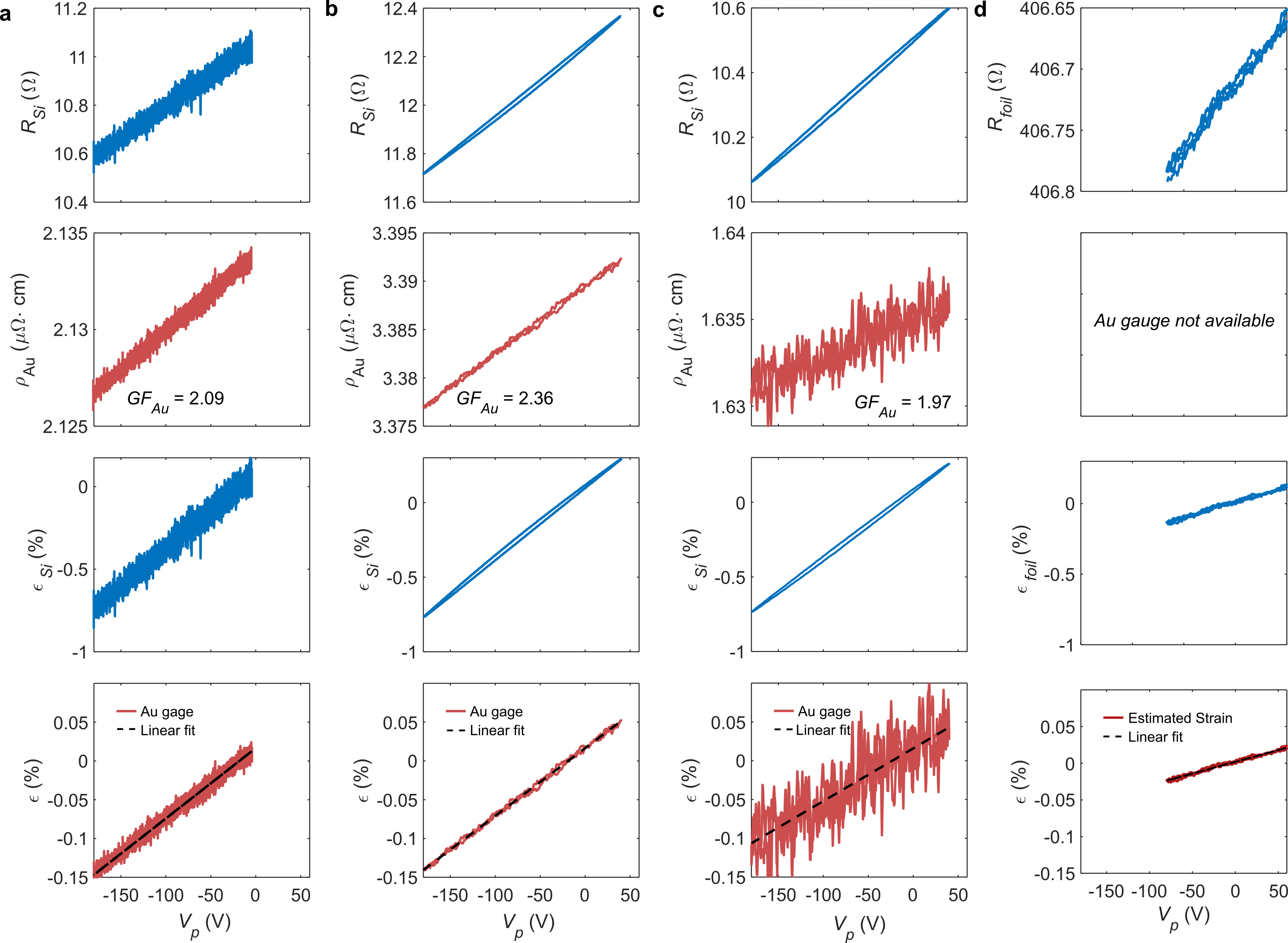}
  \caption{\textbf{Conversion of measured strain gauge resistance to $\epsilon$ for each device.}
  \textbf{a}, Data for the Bernal bilayer graphene device. The top panel shows the measured resistance of the commercial silicon strain gauge, $R_{Si}$, mounted onto one of the piezos of the strain cell. The second panel from the top shows the measured resistivity of the evaporated gold strain gauge on the Si/SiO$_2$ wafer, $\rho_{Au}$, as $V_p$ is swept back and forth over its full range. The gauge factor is calculated based on the measured value of $\rho_{Au}$ following the conversion curve shown in Ref.~\onlinecite{Liu2024StrainControl}. The third panel from the top shows the measured resistance of the commercial strain gauge converted to uniaxial strain, $\epsilon_{Si}$, using the known gauge factor of the strain gauge. The bottom panel shows the measured resistivity of the gold strain gauge converted to $\epsilon$. A linear fit to this data establishes the conversion between $V_p$ and $\epsilon$. Notably, $\epsilon$ is much smaller than $\epsilon_{Si}$, indicating imperfect strain transfer from the strain cell to the Si/SiO$_2$ substrate.
  \textbf{b}, Same measurements for the $\theta=1.10^{\circ}$ TBG device.
  \textbf{c}, Same measurements for the $\theta=1.20^{\circ}$ TBG device.
  \textbf{d}, Same measurements for the $\theta=1.31^{\circ}$ TBG device. In this device, we use a commercial foil gauge rather than a commercial silicon gauge. Owing to lack of functional gold gauge, the estimated $\epsilon$ in this device is calculated using the average of the strain transmission ratio from the other three devices (0.176).}
  \label{fig:strain_conversion}
\end{figure*}

\begin{figure*}[h]
  \centering
  \includegraphics[width=0.9\textwidth]{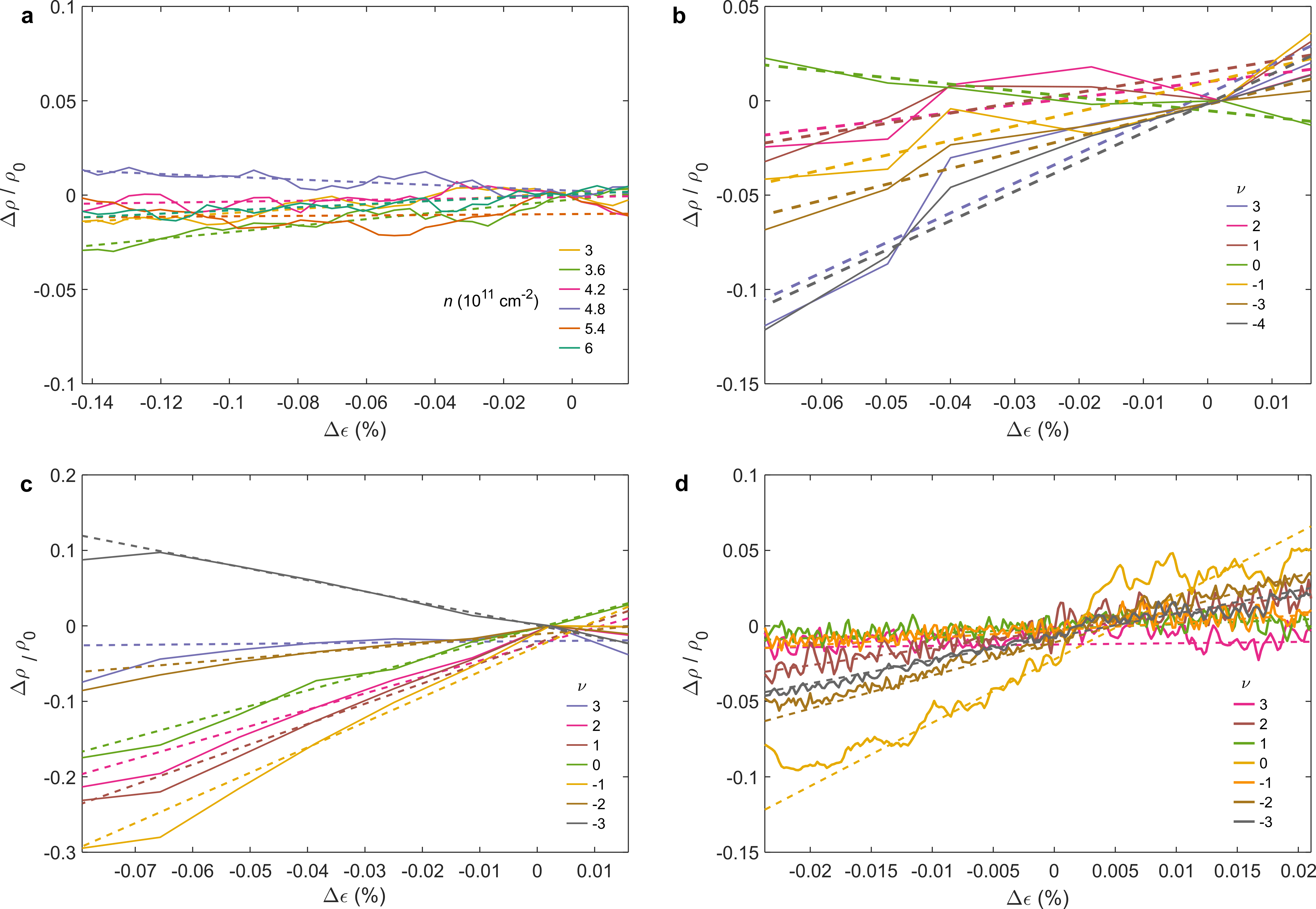}
  \caption{\textbf{Measurements of elastoresistance versus strain for all devices.} 
  \textbf{a}, Measurements for the Bernal bilayer graphene device taken at $D=0.1$~V/nm for several values of $n$. The dashed lines show linear fits to each measurement.
  \textbf{b}, The same for the $\theta=1.10^{\circ}$ TBG device for several values of $\nu$.
  \textbf{c}, The same for the $\theta=1.20^{\circ}$ TBG device.
  \textbf{d}, The same for the $\theta=1.31^{\circ}$ TBG device.}
  \label{fig:ER_Vp}
\end{figure*}

\begin{figure*}[h]
  \centering
  \includegraphics[width=0.9\textwidth]{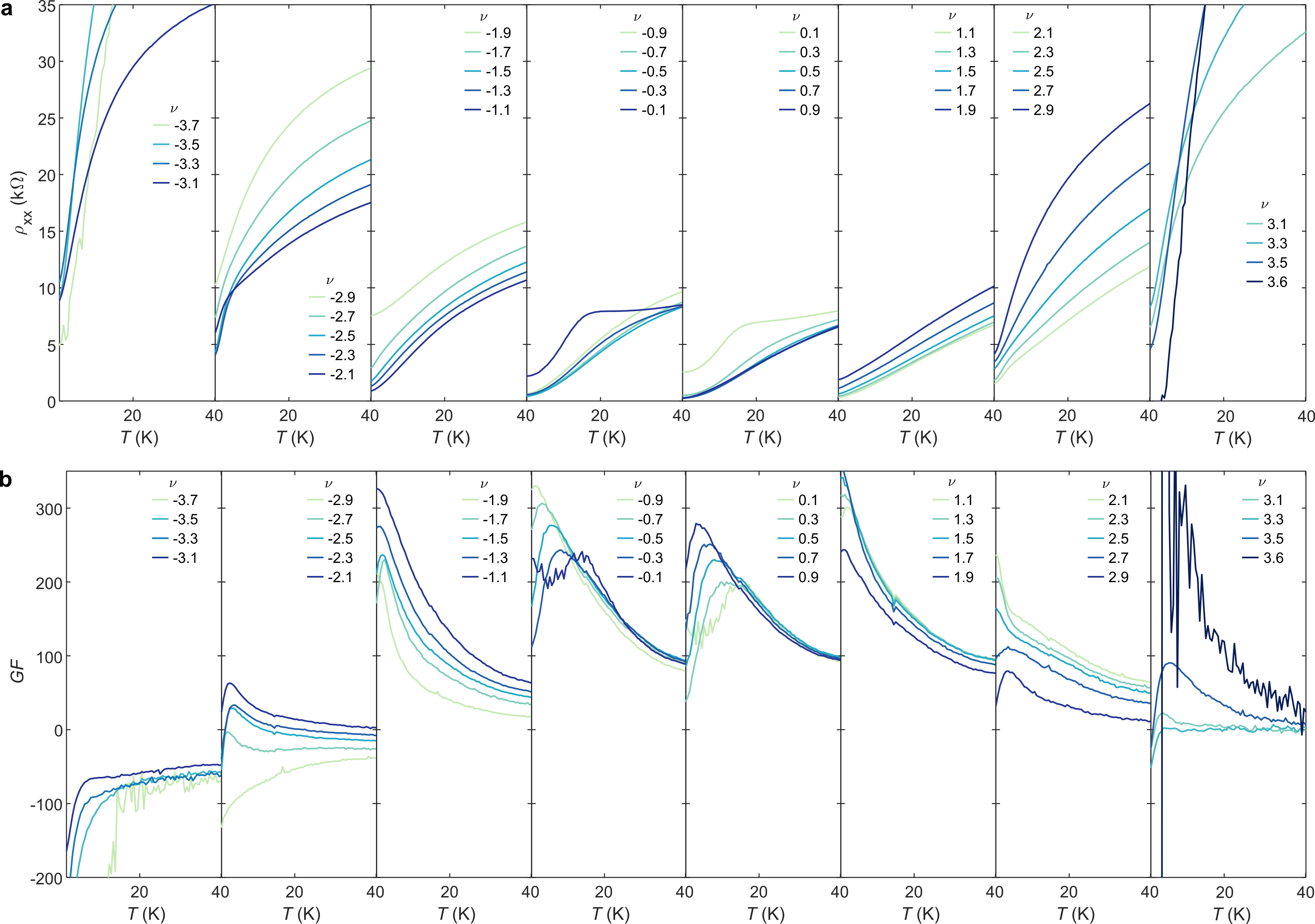}
  \caption{\textbf{Measurements of resistivity and gauge factor across the flat band for the $\theta=1.20^{\circ}$ TBG device.}
  (a) $\rho_{xx}$ versus $T$ for different $\nu$. 
  (b) $GF$ versus $T$ for the same $\nu$.}
  \label{fig:fullGF_120}
\end{figure*}

\begin{figure*}[h]
  \centering
  \includegraphics[width=0.9\textwidth]{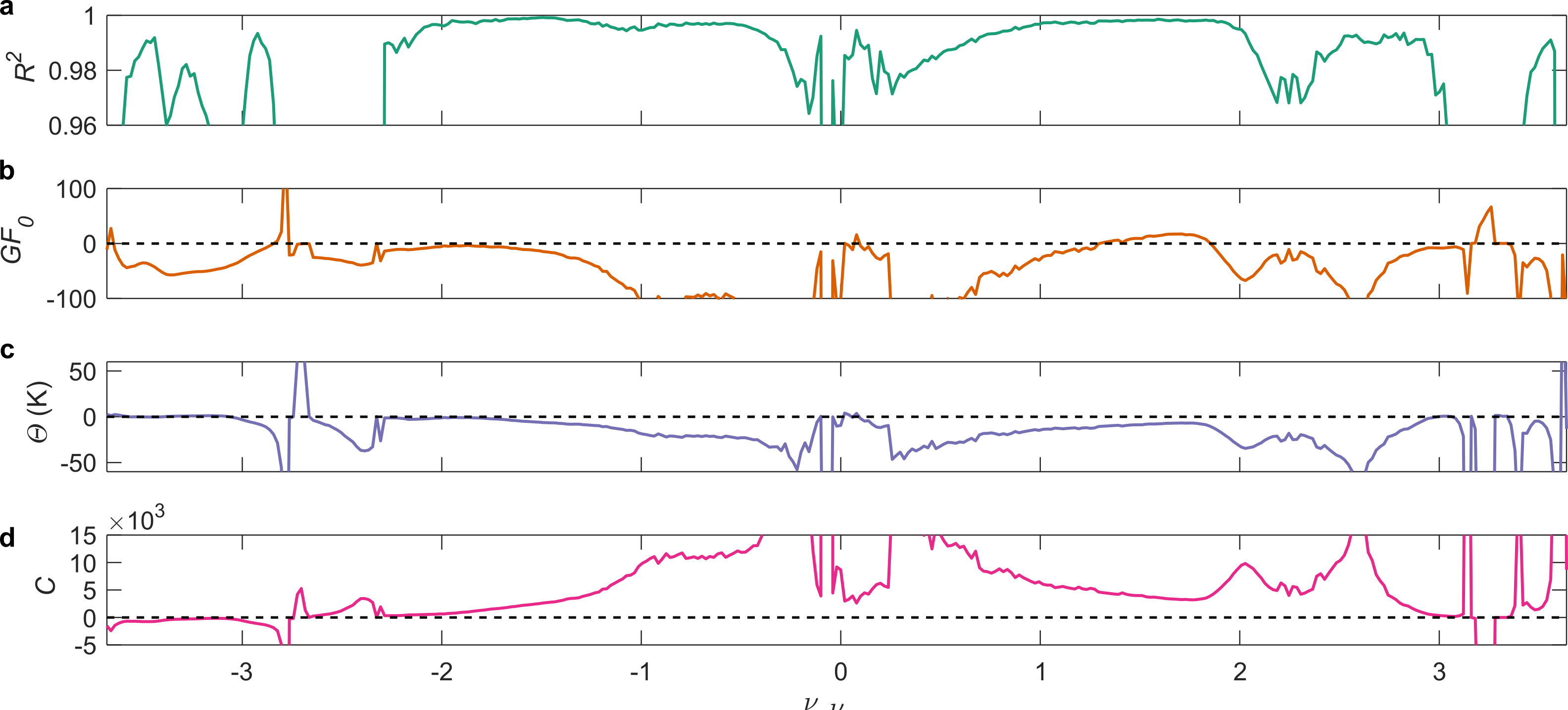}
  \caption{\textbf{Metrics of the Curie-Weiss fit for the $\theta=1.20^{\circ}$ TBG device.}
  \textbf{a}, $R^2$ value versus $\nu$ for the best fit.
  \textbf{b}, Corresponding $GF_0$.  
  \textbf{c}, Weiss temperature, $\theta$. 
  \textbf{d}, Curie constant, $C$.}
  \label{fig:CW_fit}
\end{figure*}

\begin{figure*}[h]
  \centering
  \includegraphics[width=0.9\textwidth]{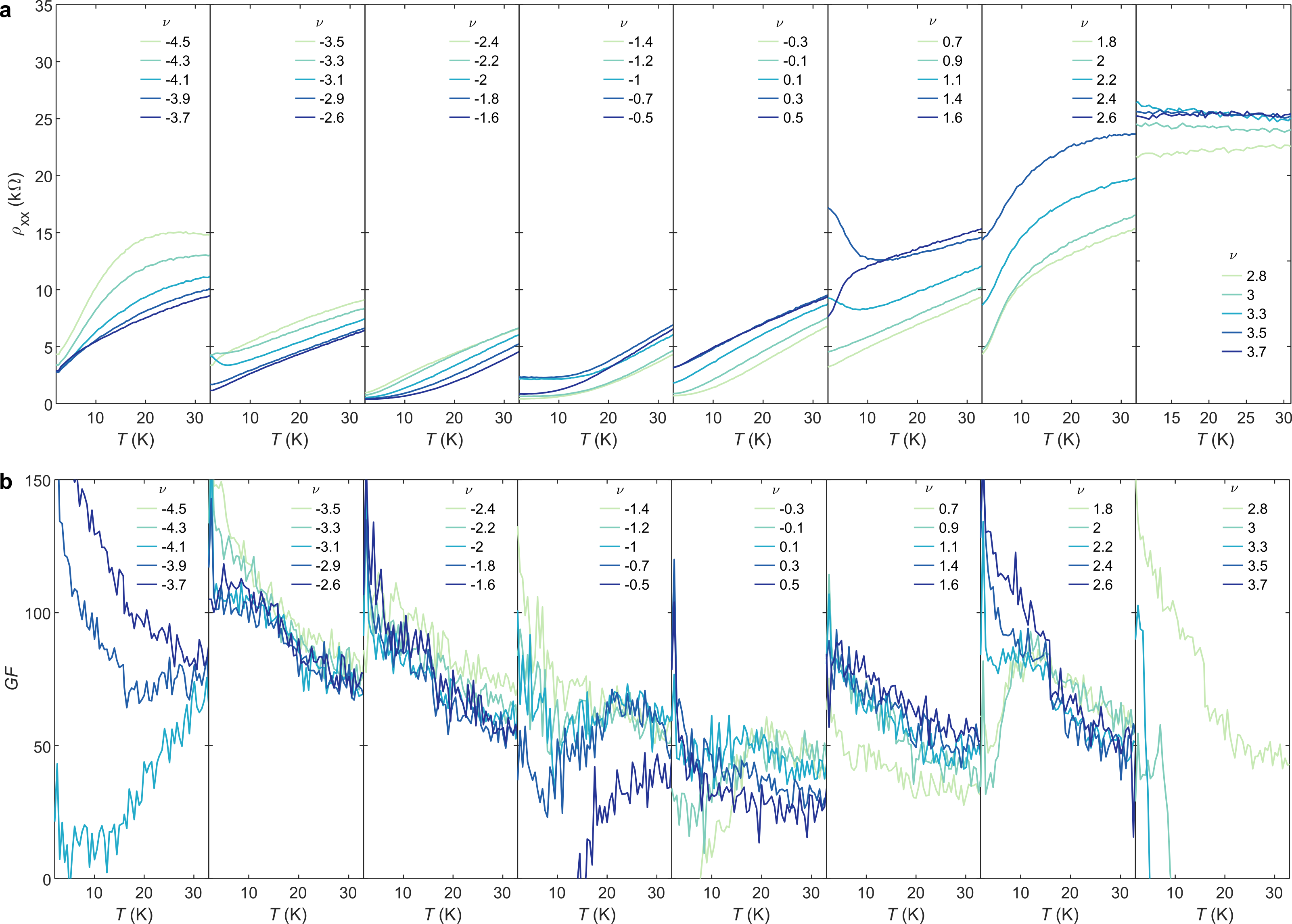}
  \caption{\textbf{Measurements of resistivity and gauge factor across the flat band for the $\theta=1.10^{\circ}$ TBG device.}
  (a) $\rho_{xx}$ versus $T$ for different $\nu$. 
  (b) $GF$ versus $T$ for the same $\nu$.}
  \label{fig:fullGF_110}
\end{figure*}

\begin{figure*}[h]
  \centering
  \includegraphics[width=0.9\textwidth]{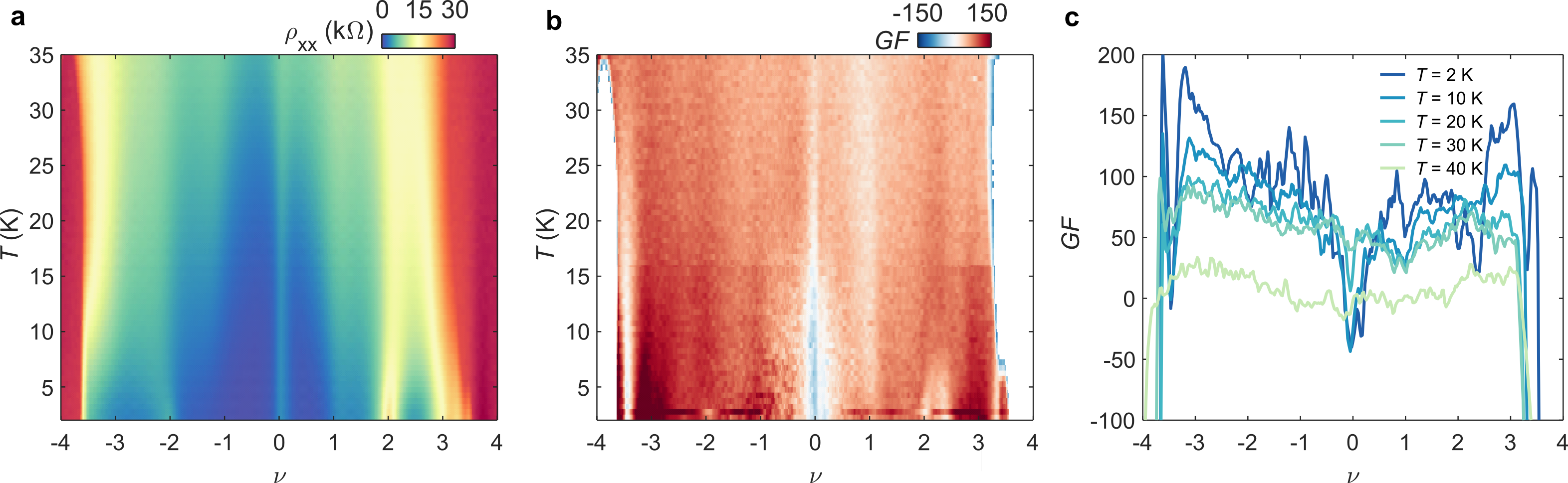}
    \caption{\textbf{Temperature dependence of the resistance and elastoresistance for the $\theta=1.10^{\circ}$ TBG device.}
    \textbf{a}, Measurement of $\rho_{xx}$ versus $\nu$ and $T$.
    \textbf{b}, Map of the $GF$ versus $\nu$ and $T$, acquired in the same way as Fig.~3b of the main text.
    \textbf{c}, Line traces of $GF$ versus $\nu$ for several fixed $T$.}
  \label{fig:GFtemp_110}
\end{figure*}

\begin{figure*}[h]
  \centering
  \includegraphics[width=\textwidth]{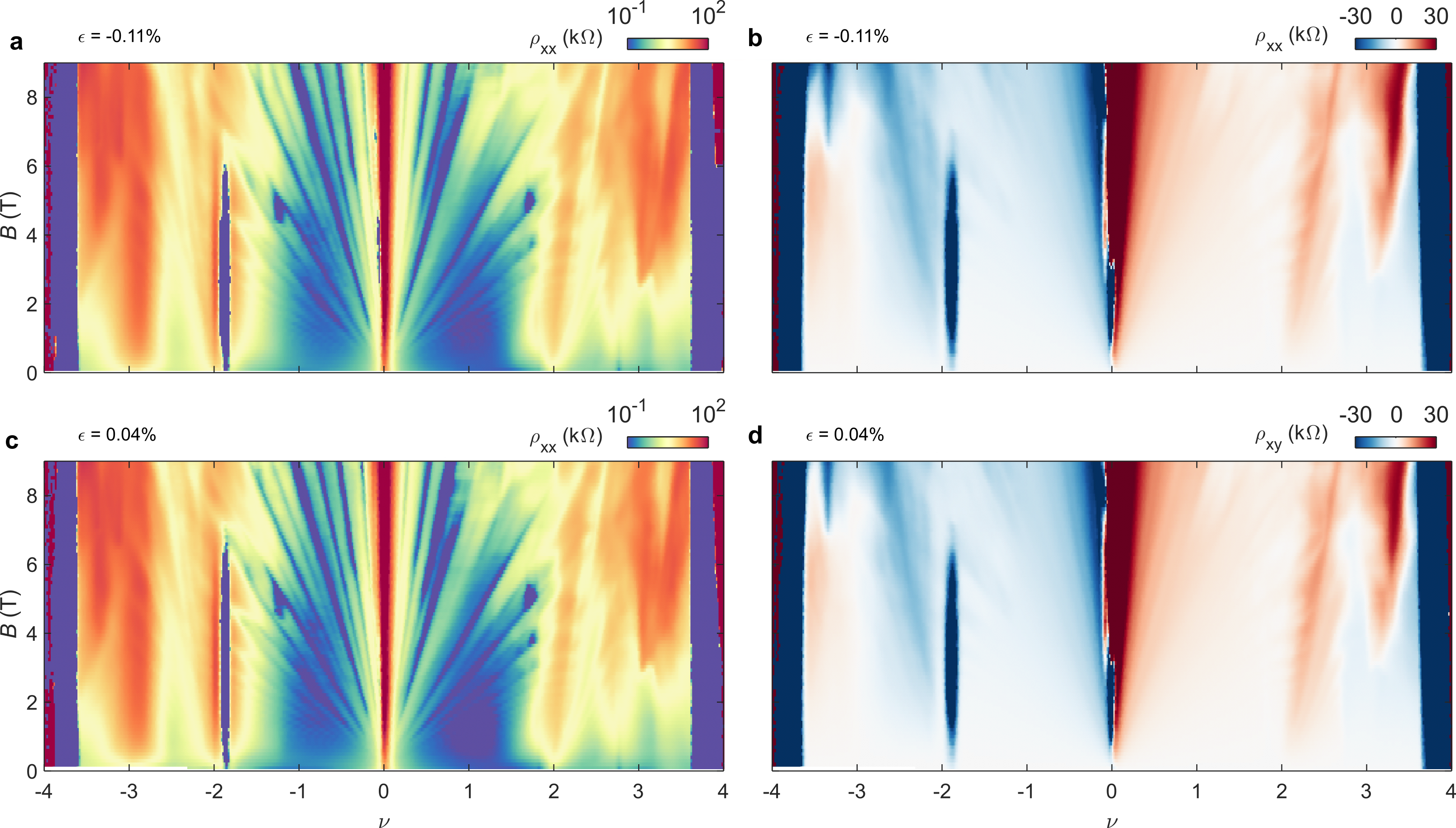}
  \caption{\textbf{Landau fan diagrams at two values of $\epsilon$ from the $\theta=1.20^{\circ}$ TBG device.}
  \textbf{a-b}, Landau fans $\rho_{xx}$ (left) and $\rho_{xy}$ (right) at $\epsilon = -0.11\%$;
  \textbf{c-d}, The same measurements at $\epsilon = 0.04\%$. In both measurements, the same conversion is used from gate voltage to $\nu$. Within experimental resolution, all quantum oscillations project to the same values of $\nu$ irrespective of $\epsilon$. All data are taken at $T=100$~mK.}
  \label{fig:fans_strain}
\end{figure*}

\begin{figure*}[h]
  \centering
  \includegraphics[width=0.9\textwidth]{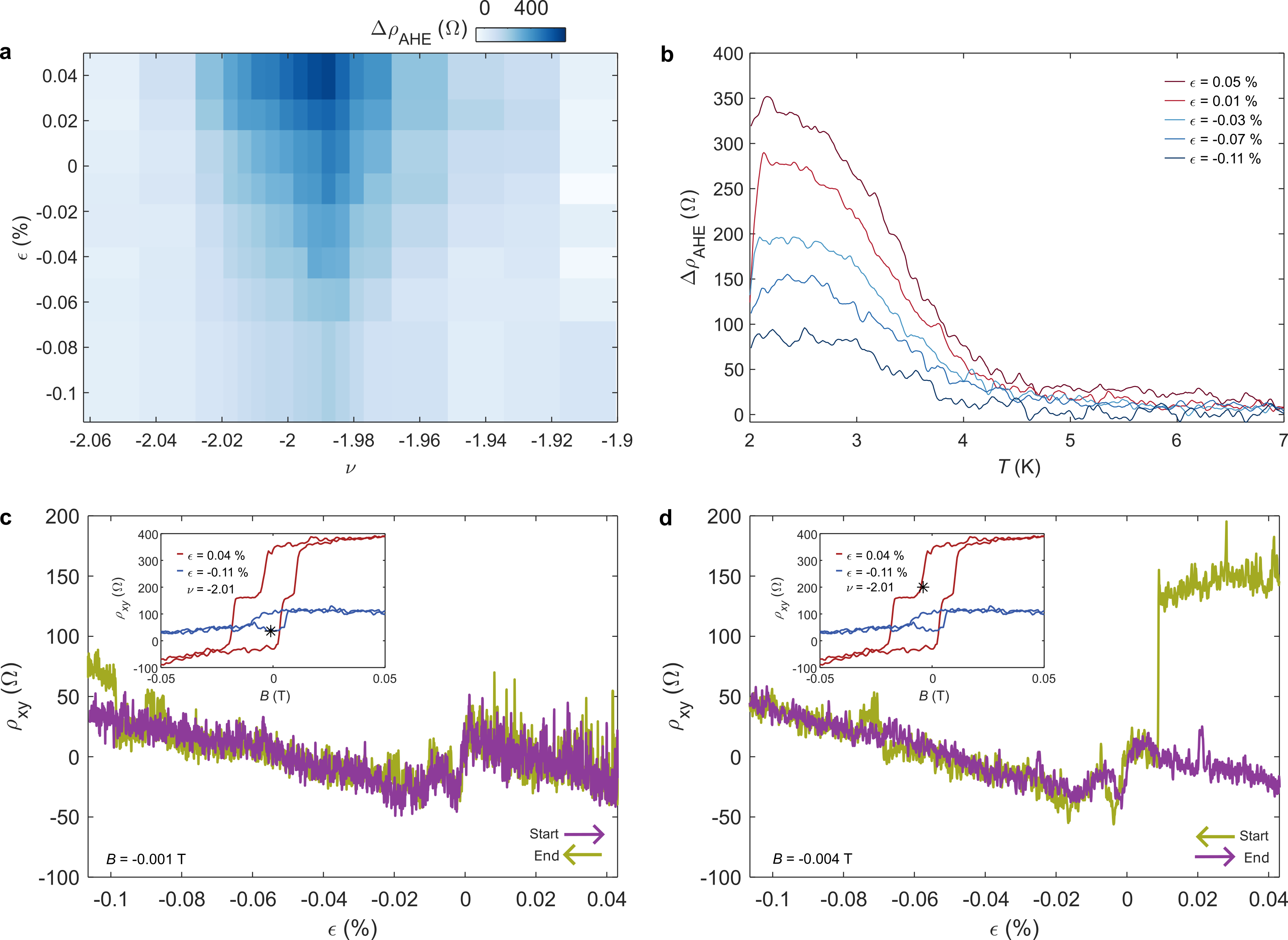}
  \caption{\textbf{Dependence of the AHE on strain in the $\theta=1.20^{\circ}$ TBG device.}
  (a) Map of the anomalous Hall amplitude, $\Delta \rho_{\mathrm{AHE}}$ versus $\nu$ and $\epsilon$. $\Delta \rho_{\mathrm{AHE}}$ is calculated as the difference between the maximum and minimum values of $\rho_{xy}$ in the the anomalous Hall loop.
  (b) Measurement of $\Delta \rho_{\mathrm{AHE}}$ versus $T$ for several $\epsilon$.
  (c) Measurement of $\rho_{xy}$ as $V_p$ is swept back and forth at $B=-0.001$~T at 100 mK with $\nu=-2.01$. The loop first goes from compressive to tensile, then tensile to compressive. The inset shows measurements of $\rho_{xy}$ versus $B$ at the two extremal values of $\epsilon$. The black star marker denotes the starting point of the measurement in the main panel.
  (d) The same as \textbf{c}, but at $B=-0.004$~T and starting first on the tensile side. The irreversibility likely indicates magnetic domain wall motion.}
  \label{fig:AHE_strain}
\end{figure*}

\end{document}